\documentclass[12pt]{article}
\usepackage{graphicx,bm,amsmath,amssymb,latexsym,psfrag,epsfig,subfigure,euscript, multirow,color,verbatim}
\usepackage{bbm}
\usepackage{axodraw}
\usepackage{slashed}

\allowdisplaybreaks

\newcommand{\ket}[1]{\left| #1 \right>} % for Dirac bras
\newcommand{\bra}[1]{\left< #1 \right|} % for Dirac kets
\newcommand{\e}{\epsilon}

\newcommand{\non}{\nonumber}
\newcommand{\nn}{\nonumber \\}

\newcommand{\nbar}{{\bar{n}}}

\def\rd{\mathrm{d}}

\def\rd{\mathrm{d}}

\definecolor{darkred}{rgb}{0.7,0.0,0.0}
\definecolor{darkblue}{rgb}{0.0,0.0,0.9}
\definecolor{darkgreen}{rgb}{0.0,0.5,0.0}
\definecolor{brown}{rgb}{0.0,0.0,0.0}

\newcommand{\blue}{\color{darkblue}}

\newcommand{\cone}{ { c^S_1} }
\newcommand{\ctwo}{ { c^S_2} }
\newcommand{\ctwoL}{ { c^S_{2L}} }

\newcommand{\ctwor}{ { c^S_{2\rho}} }

\newcommand{\cR}{ {\mathcal{R} }}

\newcommand{\smu}{\widetilde{s}_{\mu}}
\renewcommand{\sf}{\widetilde{s}_{f}}

\newcommand{\da}{{\blue A}}
\newcommand{\db}{{\blue B}}
\newcommand{\dc}{{\blue C}}
\newcommand{\dd}{{\blue  D}}
\newcommand{\de}{{\blue E}}
\newcommand{\df}{{\blue F}}
\newcommand{\dg}{{\blue G}}
\renewcommand{\dh}{{\blue H}}
\newcommand{\di}{{\blue I}}
\renewcommand{\dj}{{\blue J}}
\newcommand{\dk}{{\blue K}}
\newcommand{\dl}{{\blue L}}
\newcommand{\dV}{{\blue \dk}}

\begin{document}
\begin{titlepage}

\begin{flushright}
\end{flushright}

\vspace{0.2cm}
\begin{center}
\Large\bf
The two-loop hemisphere soft function\\
\end{center}

\vspace{0.2cm}
\begin{center}
{\sc Randall Kelley and Matthew D. Schwartz}\\
\vspace{0.4cm}
{\sl Center for the Fundamental Laws of Nature \\
Harvard University\\
Cambridge, MA 02138, USA} \\
\vspace{0.4cm}
\vspace{0.4cm}
{\sc Robert M. Schabinger }\\
\vspace{0.4cm}

{\sl Instituto de F\'{i}sica Te\'{o}rica UAM/CSIC \\
	Universidad Aut\'{o}noma de Madrid \\
        Cantoblanco, E-28049 Madrid, Espa\~{n}a}\\
\vspace{0.4cm}
\vspace{0.4cm}
{\sc Hua Xing Zhu}\\
\vspace{0.4cm}
{\sl Department of Physics and State Key Laboratory of Nuclear Physics and Technology\\
  Peking University\\
  Beijing 100871, China} \\
\end{center}

\vspace{0.2cm}
\begin{abstract}\vspace{0.2cm}
\noindent 
The hemisphere soft function is calculated to order $\alpha_s^2$. This is the first multi-scale
soft function calculated to two loops. 
The renormalization scale dependence
of the result agrees exactly with the prediction from effective field theory. 
This fixes
the unknown coefficients of the singular parts of the two-loop thrust and heavy-jet mass distributions. There
are four such coefficients, for 2 event shapes and 2 color structures, which are shown to be
in excellent agreement with previous numerical extraction. The asymptotic behavior
of the soft function has double logs in the $C_FC_A$ color structure, which agree with non-global log calculations,
but also has sub-leading single logs for both the $C_FC_A$ and $C_F T_F n_f$ color structures. 
The general form of the soft function is complicated, does not factorize
in a simple way, and disagrees with the Hoang-Kluth ansatz. The exact hemisphere soft function will remove one
source of uncertainty on the $\alpha_s$ fits from $e^+e^-$ event shapes. 
\end{abstract}
\vfil
\end{titlepage}

%%%%%%%%%%%%%%%%%%%%%%%%%%%%%%%%%%%%%%%%%%%%%%%%%%%%%%%%%%%%%%%%%%%%%%%%%%%%%%%%%
%%%%%%%%%%%%%%%%%%%%%%%%%%%%%%%%%%%%%%%%%%%%%%%%%%%%%%%%%%%%%%%%%%%%%%%%%%%%%%%%%
%%%%%%%%%%%%%%%%%%%%%%%%%%%%%%%%%%%%%%%%%%%%%%%%%%%%%%%%%%%%%%%%%%%%%%%%%%%%%%%%%
%%%%%%%%%%%%%%%%%%%%%%%%%%%%%%%%%%%%%%%%%%%%%%%%%%%%%%%%%%%%%%%%%%%%%%%%%%%%%%%%%

\section{Introduction} \label{sec:intro}
There has been significant activity in the last few years in the effective field community to perform
accurate calculations of event shapes for $e^+ e^-$ colliders.
%There has been significant activity in the last few years in the effective field community to perform
%accurate resummed calculations of jet-based observables in high energy collisions. At hadron colliders,
%observables can be extremely complicated, with many cuts, and many relevant scales, thus it has
%been found useful to attempt to understand many of the relevant issues in the simpler context of
%inclusive hadronic event shapes at $e^+ e^-$ colliders.   
At high energy, the hadronic final states in $e^+ e^-$
collisions are dominated by the formation of jets of particles and are described by perturbative QCD.  
Comparison of theoretic calculations of event shapes with the experimentally measured values has
lead to some of the most precise measurements of the strong coupling constant $\alpha_s$.
The NNLO fixed order calculations in 
\cite{GehrmannDeRidder:2005cm, GehrmannDeRidder:2007bj, GehrmannDeRidder:2007hr, Weinzierl:2008iv}
allow the prediction of
many events shapes to order $\alpha_s^3$.   Advances in Soft-Collinear Effective Theory (SCET) 
\cite{Bauer:2000yr, Bauer:2001yt, Beneke:2002ph, Fleming:2007qr}
have allowed for resummation of large logarithmic corrections to thrust 
\cite{Schwartz:2007ib, Becher:2006qw, Becher:2008cf} and heavy jet mass
\cite{Chien:2010kc} 
 to ${\rm N^3LL}$ accuracy and non-perturbative considerations were included for thrust in 
\cite{Abbate:2010xh}. 
These results have been used to extract a value of $\alpha_s$ that is competitive with the world average
\cite{Yao:2006px}.

Dijet event shapes such as thrust and heavy jet mass demonstrate singular behavior when calculated perturbatively
at fixed order due to the appearance of large logarithmic corrections.  These large logarithms invalidate a naive
expansion in $\alpha_s$ and thus need to be resummed to provide accurate predictions in the dijet limit.
In the dijet limit, there is a clear separation between scales. 
Effective theory techniques rely on a separation between kinematic scales 
and, through renormalization group (RG) evolution, logarithms of the ratio of these scales can be resummed.  
Each of the relevant scales is described by different physics, each of which can be calculated using a different
theory.  The contribution from each can be shown to factorize into a hard contribution, due to physics at the 
center of mass energy $Q$, a jet function, due to physics at the jet scale, and a soft function which describes
soft gluon emission.  The hard and jet functions are known to 2-loops. However, the soft function relevant for 
thrust or heavy jet mass is only partially known beyond 1-loop \cite{Hoang:2008fs, Chien:2010kc}. 
In this paper, the perturbative soft function is computed analytically to order $\alpha_s^2$.

Soft functions have been studied for many years, not just in SCET. These soft functions are defined as matrix
elements of Wilson lines. For resummation up to the next-to-leading logarithmic order (NLL), all that is needed
about the soft function is its anomalous dimension.
This can be extracted either from renormalization-group invariance or from the virtual graphs. For example, such calculations
have been done for thrust~\cite{Becher:2008cf}, direct photon~\cite{Becher:2009th}, and
dijet production~\cite{Kidonakis:1997gm,Aybat:2006mz,Kelley:2010qs,Kelley:2010fn}.  To go beyond
NLL, one needs the finite parts of these soft functions, which are more difficult to calculate because the real
emission graphs are needed, and these involve often complicated phase-space cuts.
In all cases calculated at 2-loops so far,
such as Drell-Yan~\cite{Korchemsky:1993uz,Belitsky:1998tc} or $b\to s \gamma$~\cite{Becher:2005pd},
the real emission graphs only involve one scale.  Multi-scale soft functions, where different constraints
are placed on gluons or quarks going in different directions, such as the hemisphere soft function,
are likely to play an important role in hadron collisions~\cite{Ellis:2010rw, Kelley:2011tj}. At order $\alpha_s$, the multiple
scales are irrelevant, since only one gluon can be emitted. At order $\alpha_s^2$ or beyond, there can be real emission
graphs depending on multiple scales at the same time. It has been suggested~\cite{Hoang:2008fs} that the
soft function should depend only on logarithms of these scales, such as $\ln^2(k_L/k_R)$.
Whether more complicated scale-independent terms, such as  $\text{Li}_2(-{k_L/k_R}) + \text{Li}_2(-{k_R/k_L})$
might appear has been an open question. Understanding the form of these
soft functions in more detail will be important for LHC precision jet physics at NNLL and beyond~\cite{Kelley:2011tj}.

The hemisphere soft function $S(k_L,k_R,\mu)$ is the probability to have soft radiation with small component
$k_L$ going into the left hemisphere and soft radiation with small component $k_R$ going into the right hemisphere. 
More precisely, in $e^+e^- \to $ hadron events 
at center-of-mass energy $Q$, in the limit that all radiation is much softer than $Q$,
the cross section is given by matrix elements of Wilson lines. These Wilson lines point in the direction of two
back-to-back light-like quarks which come from the Born process $e^+e^- \to \bar{q} q$. Each quark direction defines a
hemisphere, which we call left and right and denote with the light-like 4-vectors  $n^\mu$ and $\bar{n}^\mu$. If the
total radiation in the left (right) hemisphere is $P_L^\mu$ ($P_R^\mu$), then $S(k_L,k_R,\mu)$ is the matrix element
squared to have $k_L= n \cdot P_L$ and $k_R = \bar{n} \cdot P_R$, with all other degrees of freedom integrated over.

The hemisphere soft function is known to have many interesting properties and is conjectured to have others. 
The factorization
theorem for the full hemisphere mass distribution implies that the Laplace transform of the soft function should factorize
into the form
%-----------------------------------------------------------------------------
\begin{equation}
\label{softfact}
\tilde{s}(L_1, L_2, \mu ) = \smu(L_1) \smu(L_2) \sf(L_1 - L_2)
\end{equation}
%-----------------------------------------------------------------------------
where $L_1 = \ln x_L \mu$ and $L_2 = \ln x_R \mu$, with $x_R$ and $x_L$ the Laplace conjugate variables to $k_L$ and $k_R$.
The anomalous dimension of the soft function and the function $\tilde{s}_\mu(L)$ are known exactly to 3-loop order. The function
$\tilde{s}_{f}(L)$ is known exactly  only to order $\alpha_s$.
Hoang and Kluth~\cite{Hoang:2008fs} argued that at order $\alpha_s^2$ the function
$\tilde{s}_{f}(L)$ must be a polynomial of at most 2nd order in $L$, i.e. $\sf(L) = c_{2}^S + c_{2L}^S L^2$. In this paper,
we show that this Hoang-Kluth ansatz does not hold; $\sf(L)$ is much more complicated.
Certain moments of $\sf(L)$ contribute to the coefficients
of $\delta(\tau)$ and $\delta(\rho)$ in the thrust and heavy-jet mass distributions. These moments were fit numerically 
in~\cite{Hoang:2008fs} and \cite{Chien:2010kc} using  numerical calculations of the singular behaviour of these distributions in full QCD with the program {\sc event 2}. In this paper, we produce these moments analytically and find that they are in excellent agreement with the most accurate available numerical fit~\cite{Chien:2010kc}.

Any $L$ dependence at large $L$ in $\sf(L)$ turns into large logarithmic behavior of the hemisphere mass distribution 
({\it i.e.} $\ln(M_L/M_R$)). Since all
of the $\mu$ dependence is in $\smu(L)$, these large logs are not determined by RG invariance and
correspond to so-called ``non-global logs''. Dasgupta and Salam
calculated the non-global logs for the related left-hemisphere mass distribution in full QCD~\cite{Dasgupta:2001sh} and found 
no non-global logs (up to order $L^2$) for the $C_F n_f T_F$ color structure and an $L^2$ term with coefficient $-\frac{4\pi^2}{3}$ 
for the $C_F C_A$ term.
% This result was reproduced, also assuming stronly-ordered emissions, starting from the soft limit of QCD in~\cite{scetCL}.
We show below that the asymptotic behavior of $\sf(L)$ in the full soft function
is indeed of the form $-\frac{4\pi^2}{3} L^2$ for the $C_F C_A$ color structure.
We also find that both this color structure and the $C_F n_f T_F$ one have additional non-global single logs. These are especially interesting
because the soft function is symmetric in $L \to -L$, which seems to forbid a linear term. 
The linear term appears through a complicated analytic function involving polylogarithms which actually asymptotes to $|L|$.

This paper is organized as follows.
 In section~\ref{sec:shapes} we review the factorization formula for the hemisphere mass distribution and its thrust and heavy-jet mass projections. 
Section~\ref{sec:calculation} computes the soft function in dimensional regularization. The calculation
is complicated, so the results are summarized separately in~\ref{sec:summary}.
Section~\ref{sec:integrating} 
discusses the result and presents the renormalized result for the integrated
soft function, which can be compared directly to the predictions from SCET. Section
~\ref{sec:thrust} gives the previously missing terms in the singular parts of
the 2-loop thrust and heavy jet mass distributions, and compares to previous numerical estimates.
Section~\ref{sec:hemi} gives the full integrated hemisphere soft function which is 
compared to previous conjectures. The asymptotic form of this distribution, which exhibits non-global
logs, is discussed in Section~\ref{sec:asym}. Section~\ref{sec:exp} has some comments on predicting
higher order terms with non-Abelian exponentiation.
Conclusions and implications are discussed in Section~\ref{sec:conc}.

%%%%%%%%%%%%%%%%%%%%%%%%%%%%%%%%%%%%%%%%%%%%%%%%%%%%%%%%%%%%%%%%%%%%%%%%%%%%%%%%%
%%%%%%%%%%%%%%%%%%%%%%%%%%%%%%%%%%%%%%%%%%%%%%%%%%%%%%%%%%%%%%%%%%%%%%%%%%%%%%%%%
%%%%%%%%%%%%%%%%%%%%%%%%%%%%%%%%%%%%%%%%%%%%%%%%%%%%%%%%%%%%%%%%%%%%%%%%%%%%%%%%%
%%%%%%%%%%%%%%%%%%%%%%%%%%%%%%%%%%%%%%%%%%%%%%%%%%%%%%%%%%%%%%%%%%%%%%%%%%%%%%%%%

\section{Event Shapes and Factorization in SCET \label{sec:shapes}}
The hemisphere soft function appears in the factorization theorem for the hemisphere mass distribution. 
The hemispheres are defined
with respect to the thrust axis. Thrust itself is defined by
%-----------------------------------------------------------------------------
\begin{equation}
T = 
  \max_{\mathbf n}
  \left(
  \frac
  {
    \sum_{i} | {\mathbf p}_i \cdot {\mathbf{n}}  | 
   }
   {
    \sum_{i} | \mathbf{p}_i | 
   }
   \right), 
\end{equation}
%-----------------------------------------------------------------------------
where the sum is over all momentum 3-vectors $\mathbf{p}_i$ in the event.  The thrust axis is the
unit 3-vector $\mathbf{n}$ that maximizes the expression in parentheses. We then define the light-like 4-vectors
$n^\mu = (1,\mathbf{n})$ and $\bar{n}^\mu = (1,-\mathbf{n})$. In the dijet limit $T \to 1$ and it is therefore more convenient to define $\tau = 1 - T$ as the thrust variable so that $\tau$ is small in the dijet limit.  

Once the thrust axis is known, we divide the event into two hemispheres defined by the plane perpendicular 
to the thrust axis.  We define $P_L^{\mu}$ and $P_R^{\mu}$ to be the 4-vector sum of all of the radiation 
going into each hemisphere and $M_L = \sqrt{P_L^2}$ and $M_R = \sqrt{P_R^2}$ to be the hemisphere invariant masses.
When both $M_{L}$ and $M_{R}$ are small compared to the center-of-mass energy, $Q$, the hemisphere mass distribution
factorizes into~\cite{Fleming:2007qr}
%-----------------------------------------------------------------------------
\begin{equation}
\label{FacTheorem}
\frac{1}{\sigma_0}\frac{\rd^2 \sigma}{\rd M_L^2 \rd M_R^2} 
=  H(Q^2, \mu) 
  \int \rd k_L \rd k_R
  J( M_L^2 - Qk_L, \mu) J( M_R^2 - Qk_R, \mu)
  S(k_L, k_R, \mu)\,.
\end{equation}
%-----------------------------------------------------------------------------
%The subscript on $\sigma_2$ is a reminder that his formula holds only in the dijet limit (it can be matched
%to full QCD to give a result valid in general).
Here, $\sigma_0$ is the tree level total cross section. $H(Q^2, \mu)$ is the hard function
which accounts for the matching between QCD and SCET.  $J(p^2)$ is the inclusive jet function which accounts
for the matching between an effective field theory with soft and collinear modes to a theory with only soft modes.
Finally, the object of interest, $S(k_L, k_R, \mu)$ is the hemisphere soft function,
which is derived by integrating out the remaining soft modes. 

In the threshold limit (small hemisphere masses), the thrust axis aligns 
with the jet axis and thrust can be written as the sum of the two hemisphere masses,
%-----------------------------------------------------------------------------
\begin{equation}
\tau = \frac{M_L^2 + M_R^2}{Q^2} + \mathcal{O}\left( \frac{ M_{L,R}^4}{Q^4} \right)
\end{equation}
%-----------------------------------------------------------------------------
Heavy jet mass $\rho$ is defined to be the larger of the two hemisphere masses, normalized to the center of mass
energy $Q$, 
%-----------------------------------------------------------------------------
\begin{equation}
\rho = 
  \frac{1}{Q^2}
  \max( M_L^2 , M_R^2 ) .
\end{equation}
%-----------------------------------------------------------------------------
When $\rho$ is small, both hemisphere masses are small and the event 
appears as two pencil-like, back to back jets.

The factorization formula can be used to calculate thrust and heavy jet mass
in the dijet limit as integrals over the doubly differential hemisphere mass distribution.
Explicitly,
%-----------------------------------------------------------------------------
\begin{equation}
\label{def:thrust}
\frac{\rd \sigma}{\rd \tau} 
  = Q^2 \int \rd M_L^2 \rd M_R^2
  \frac{\rd^2 \sigma}{\rd M_L^2 \rd M_R^2} 
  \delta (Q^2 \tau - M_L^2 - M_R^2 ) 
\end{equation}
%-----------------------------------------------------------------------------
and
%-----------------------------------------------------------------------------
\begin{equation}
\label{def:hjm}
\frac{\rd \sigma}{\rd \rho} 
  = Q^2 \int \rd M_L^2 \rd M_R^2
  \frac{\rd^2 \sigma}{\rd M_L^2 \rd M_R^2} 
  \left[
    \delta (Q^2 \rho - M_L^2)\theta( M_L^2 - M_R^2)  
    +
    \delta (Q^2 \rho - M_R^2)\theta( M_R^2 - M_L^2)  
  \right]\,.
\end{equation}
%-----------------------------------------------------------------------------

The thrust distribution can be written so that it depends not on the full hemisphere soft function
but on the thrust-soft function, defined as
\begin{equation}
S_T(k,\mu) = \int \rd k_L \rd k_R  S(k_L ,k_R,\mu) \delta(k-k_L -k_R)\,.
\end{equation}
Since the thrust soft function is dimensionless and its $\mu$ dependence is determined by renormalization
group invariance, the $k$ dependence is also completely known. Thus at each order in $\alpha_s$ only one
number, the constant part, is unknown. In contrast, for the heavy jet mass distribution, the full
$k_L$ and $k_R$ dependence of the soft function is needed for the factorization theorem. In particular,
for resummation to  N$^3$LL order, only one number is needed for thrust (the constant in the 2-loop thrust soft
function), which has been fit numerically, but for heavy-jet mass a function is needed~\cite{Chien:2010kc}. In
this paper we compute both the number and the function.

%%%%%%%%%%%%%%%%%%%%%%%%%%%%%%%%%%%%%%%%%%%%%%%%%%%%%%%%%%%%%%%%%%%%%%%%%%%%%%%%%
%%%%%%%%%%%%%%%%%%%%%%%%%%%%%%%%%%%%%%%%%%%%%%%%%%%%%%%%%%%%%%%%%%%%%%%%%%%%%%%%%
%%%%%%%%%%%%%%%%%%%%%%%%%%%%%%%%%%%%%%%%%%%%%%%%%%%%%%%%%%%%%%%%%%%%%%%%%%%%%%%%%
%%%%%%%%%%%%%%%%%%%%%%%%%%%%%%%%%%%%%%%%%%%%%%%%%%%%%%%%%%%%%%%%%%%%%%%%%%%%%%%%%

%%%%%%%%%%%%%%%%%%%%%%%%%%%%%%%%%%%%%%%%%%%%%%%%%%%%%%%%%%%%%%%%%%%%%%%%%%%%%%%%%
%%%%%%%%%%%%%%%%%%%%%%%%%%%%%%%%%%%%%%%%%%%%%%%%%%%%%%%%%%%%%%%%%%%%%%%%%%%%%%%%%
%%%%%%%%%%%%%%%%%%%%%%%%%%%%%%%%%%%%%%%%%%%%%%%%%%%%%%%%%%%%%%%%%%%%%%%%%%%%%%%%%
%%%%%%%%%%%%%%%%%%%%%%%%%%%%%%%%%%%%%%%%%%%%%%%%%%%%%%%%%%%%%%%%%%%%%%%%%%%%%%%%%

\section{Calculation of the Soft Function \label{sec:calculation}}
The soft function is defined as
%-----------------------------------------------------------------------------
\begin{equation}
S(k_L, k_R, \mu) 
  \equiv 
  \frac{1}{N_c} \sum_{X_s}
  \delta( k_R - n \cdot P_s^{R} ) 
  \delta( k_L - \nbar \cdot P_s^{L} ) 
  \bra{0} \overline{Y}_{\nbar} Y_n \ket{ X_s } 
  \bra{X_s} Y^{\dagger}_n \overline{Y}^{\dagger}_{\nbar} \ket{ 0 } ,
\end{equation}
%-----------------------------------------------------------------------------
where $P_s^{L,R}$ is the total momentum of the final state $\ket{X_s}$
in the left and right hemisphere, respectively. The Wilson lines $Y_n$ and $\overline{Y}_{\nbar}$
are defined by
%-----------------------------------------------------------------------------
\begin{align}
  Y^{\dagger}_n(x) 
    = P \exp \left( ig \int_0^{\infty}\!\! ds\ n \cdot A_s(n s + x) \right) 
  &&
  \overline{Y}^{\dagger}_{\nbar}(x) 
    = P \exp \left( ig \int_0^{\infty}\!\! ds\ \nbar \cdot \overline{A}_s(n s + x) \right) ,
\end{align}
%-----------------------------------------------------------------------------
where $P$ denotes path ordering and $A_s = A^{a}_s T^a$  
$(\overline{A}_s = A_s^{a} \overline{T}^{a})$ are gauge fields in the fundamental (anti-fundamental)
representation.  The soft function can be factorized into a perturbative
(partonic) part and non-perturbative part which has support of order $\Lambda_{QCD}$~\cite{Abbate:2010xh}.

The authors of \cite{Hoang:2008fs} observed that the form of the soft function is 
constrained by the non-Abelian exponentiation
theorem and RG invariance, which puts constraints on powers of logarithms of $\mu$.
The theorem also restricts the $C_F^n$ color structure in the soft function to
be completely determined by the one-loop result. Beyond this, however, the soft function is
unconstrained.  The one-loop calculation was done in~\cite{Schwartz:2007ib,Fleming:2007xt}.
The main result of this paper is the calculation of the perturbative part of the 
hemisphere soft function to order $\alpha_s^2$. Since the order $\alpha_s^2$ color 
structure $C_F^2$ is given in \cite{Hoang:2008fs},
we will only calculate the $C_F C_A$ and the $C_F n_f T_F$ terms. 

\subsection{$C_F C_A$ color structure}
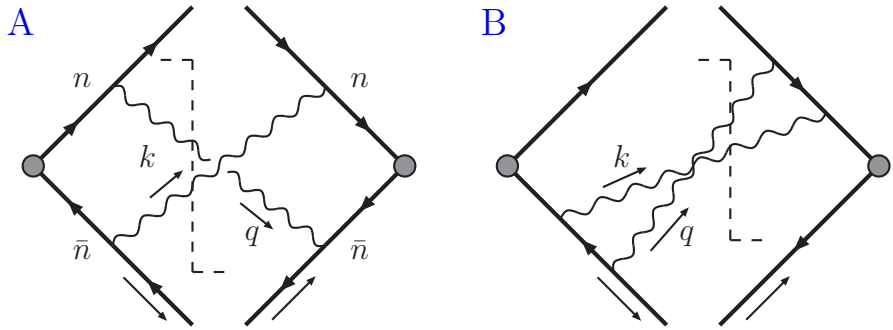
\begin{figure}[t]
\begin{center}
 \begin{picture}(200,150)(-30,0) % (-25,-30)
    \SetScale{0.4}

    \SetWidth{4}

    %fermion lines

    \ArrowLine(0,150)(75,225)
    \ArrowLine(75,225)(150,300)
    %\ArrowLine(150,300)(0,150)

    \ArrowLine(75,75)(0,150)
    \ArrowLine(150,0)(75,75)
    %\ArrowLine(0,150)(150,0)

    \ArrowLine(275,75)(200,0)
    \ArrowLine(350,150)(275,75)
    %\ArrowLine(200,0)(350,150)

    \ArrowLine(275,225)(350,150)
    \ArrowLine(200,300)(275,225)
    %\ArrowLine(350,150)(200,300)

    \SetWidth{2}

    %photon line
    \Photon(75,225)(167,156){5}{4}
    \Photon(183,144)(275,75){5}{4}
    \Photon(175,150)(275,225){5}{4}
    \Photon(75,75)(175,150){5}{4}

    %cut lines
    \DashLine(150,250)(150,50){10}
    \DashLine(120,250)(150,250){10}
    \DashLine(180,50)(150,50){10}

    %Diagram label
    \Text(-10,110)[lb]{\large{\da}}

    %momentum labels
    \LongArrow(85,45)(125,05)
    \LongArrow(225,05)(265,45)

    \LongArrow(110,120)(140,145)
    \Text(40,60)[lb]{{\Black{$k$}}}

    \LongArrow(195,115)(225,91)
    \Text(80,30)[lb]{{\Black{$q$}}}

    \Text(15,90)[lb]{{\Black{$n$}}}
    \Text(15,25)[lb]{{\Black{$\nbar$}}}
    \Text(120,90)[lb]{{\Black{$n$}}}
    \Text(120,25)[lb]{{\Black{$\nbar$}}}

    %vertex
    \SetColor{Black}
    \COval(350,150)(10,10)(10){Black}{Gray}
    \COval(0,150)(10,10)(10){Black}{Gray}
\end{picture}
\hspace{-1cm}
\begin{picture}(200,150)(-30,0) % (-25,-30)
    \SetScale{0.4}

    \SetWidth{4}

    %fermion lines
    \ArrowLine(150,0)(0,150)

    \ArrowLine(0,150)(150,300)

    \ArrowLine(350,150)(200,0)

    \ArrowLine(200,300)(350,150)

    \SetWidth{2}

    %photon line
    \Photon(50,100)(175,150){5}{4}
    \Photon(100,50)(175,150){5}{4}

    \Photon(300,200)(175,150){5}{4}
    \Photon(250,250)(175,150){5}{4}

    %cut lines
    \DashLine(210,250)(210,80){10}
    \DashLine(180,250)(210,250){10}
    \DashLine(240,80)(210,80){10}

    %Diagram label
    \Text(-10,110)[lb]{\large{\db}}

    %momentum labels
    \LongArrow(85,45)(125,05)
    \LongArrow(225,05)(265,45)

    \LongArrow(90,130)(130,148)
    \Text(40,60)[lb]{{\Black{$k$}}}

    \LongArrow(135,70)(170,110)
    \Text(65,30)[lb]{{\Black{$q$}}}

    %vertex
    \SetColor{Black}
    \COval(350,150)(10,10)(10){Black}{Gray}
    \COval(0,150)(10,10)(10){Black}{Gray}
 \end{picture}
\end{center}
\caption{Diagrams \da~and \db~contribute to both $C_F^2$ and $C_F C_A$ color factors.}
\label{fig:1}
\end{figure}

\begin{figure}[h]
\begin{center}
\begin{picture}(200,150)(-30,0)
\SetScale{0.4}

    \SetWidth{4}

    %fermion lines
    \ArrowLine(0,150)(150,300)

    \ArrowLine(75,75)(0,150)
    \ArrowLine(150,0)(75,75)

    \ArrowLine(275,75)(200,0)
    \ArrowLine(350,150)(275,75)

    \ArrowLine(275,225)(350,150)
    \ArrowLine(200,300)(275,225)

    \SetWidth{2}

    %cut lines
    \DashLine(200,250)(200,50){10}
    \DashLine(170,250)(200,250){10}
    \DashLine(230,50)(200,50){10}

    %photon line
    \Photon(175,150)(275,75){5}{4}
    \Photon(175,150)(275,225){5}{4}
    \Photon(75,75)(175,150){5}{4}

    %Diagram label
    \Text(-10,110)[lb]{\large{\dc}}
    %momentum labels
        \LongArrow(85,45)(125,05)
        \LongArrow(225,05)(265,45)

    \LongArrow(110,120)(140,145)
        \Text(25,60)[lb]{{\Black{$q+k$}}}

    \LongArrow(235,125)(265,101)
        \Text(105,47)[lb]{{\Black{$q$}}}

    \LongArrow(235,170)(265,194)
        \Text(105,64)[lb]{{\Black{$k$}}}

    %vertex
        \SetColor{Black}
    \COval(350,150)(10,10)(10){Black}{Gray}
    \COval(0,150)(10,10)(10){Black}{Gray}
\end{picture}
\hspace{-1cm}
\begin{picture}(200,150)(-30,0)
    \SetScale{0.4}

    \SetWidth{4}

    %fermion lines
    \ArrowLine(0,150)(75,225)
    \ArrowLine(75,225)(150,300)

    \ArrowLine(150,0)(0,150)

    \ArrowLine(275,75)(200,0)
    \ArrowLine(350,150)(275,75)

    \ArrowLine(275,225)(350,150)
    \ArrowLine(200,300)(275,225)

    \SetWidth{2}

    %cut lines
    \DashLine(200,250)(200,50){10}
    \DashLine(170,250)(200,250){10}
    \DashLine(230,50)(200,50){10}

    %photon line
    \Photon(75,225)(175,150){5}{4}
    \Photon(175,150)(275,75){5}{4}
    \Photon(175,150)(275,225){5}{4}

    %Diagram label
    \Text(-10,110)[lb]{\large{\dd}}

    %momentum labels
    \LongArrow(85,45)(125,05)
    \LongArrow(225,05)(265,45)

    %vertex
    \SetColor{Black}
    \COval(350,150)(10,10)(10){Black}{Gray}
    \COval(0,150)(10,10)(10){Black}{Gray}
\end{picture}
\end{center}
\caption{Diagrams \dc~and \dd~contribute to the $C_F C_A$ color factor.}
\label{fig:2}
\end{figure}
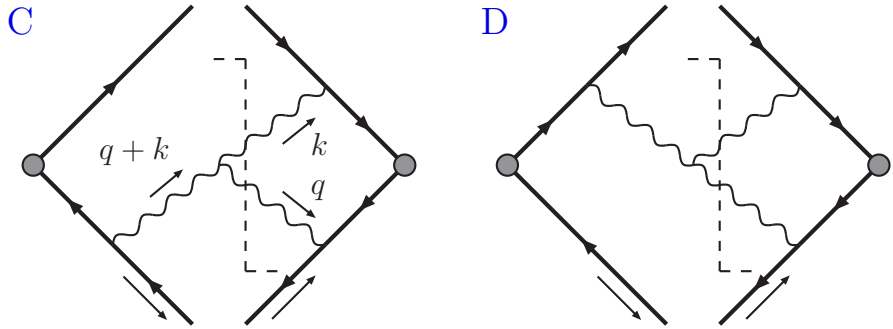

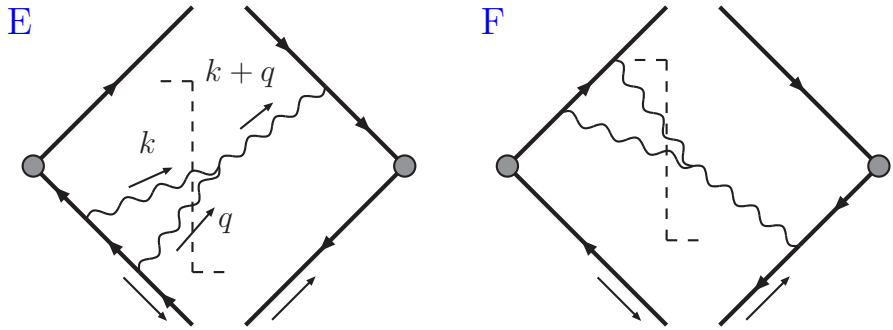
\begin{figure}
\begin{center}
\begin{picture}(200,150)(-30,0) % (-25,-30)
    \SetScale{0.4}

    \SetWidth{4}

    %fermion lines
    \ArrowLine(275,225)(350,150)
    \ArrowLine(200,300)(275,225)

    \ArrowLine(50,100)(0,150)
    \ArrowLine(100,50)(50,100)
    \ArrowLine(150,0)(100,50)

    \ArrowLine(0,150)(150,300)

    \ArrowLine(350,150)(200,0)

    \SetWidth{2}

    %photon line
    \Photon(50,100)(175,150){5}{4}
    \Photon(100,50)(175,150){5}{4}

    \Photon(175,150)(275,225){5}{4}

    %cut lines
    \DashLine(150,230)(150,50){10}
    \DashLine(120,230)(150,230){10}
    \DashLine(180,50)(150,50){10}

    %Diagram label
    \Text(-10,110)[lb]{\large{\de}}

    %momentum labels
    \LongArrow(85,45)(125,05)
    \LongArrow(225,05)(265,45)

    \LongArrow(195,185)(225,209)
    \Text(65,90)[lb]{{\Black{$k+q$}}}

    \LongArrow(90,130)(130,148)
    \Text(40,65)[lb]{{\Black{$k$}}}

    \LongArrow(135,70)(170,110)
    \Text(70,35)[lb]{{\Black{$q$}}}

    %vertex
    \SetColor{Black}
    \COval(350,150)(10,10)(10){Black}{Gray}
    \COval(0,150)(10,10)(10){Black}{Gray}
\end{picture}
\hspace{-1cm}
\begin{picture}(200,150)(-30,0) % (-25,-30)
    \SetScale{0.4}

    \SetWidth{4}

    %fermion lines
    \ArrowLine(275,75)(200,0)
    \ArrowLine(350,150)(275,75)

    \ArrowLine(200,300)(350,150)

    \ArrowLine(150,0)(0,150)

    \ArrowLine(0,150)(150,300)

    \SetWidth{2}

    %photon line
    \Photon(175,150)(275,75){5}{4}

    \Photon(100,250)(175,150){5}{4}
    \Photon(50,200)(175,150){5}{4}

    %cut lines
    \DashLine(150,250)(150,80){10}
    \DashLine(120,250)(150,250){10}
    \DashLine(180,80)(150,80){10}

    %Diagram label
    \Text(-10,110)[lb]{\large{\df}}

    %momentum labels
    \LongArrow(85,45)(125,05)
    \LongArrow(225,05)(265,45)

    %vertex
    \SetColor{Black}
    \COval(350,150)(10,10)(10){Black}{Gray}
    \COval(0,150)(10,10)(10){Black}{Gray}
\end{picture}
\end{center}
\caption{Diagrams \de~and \df~contribute to the $C_F C_A$ color factor.}
\label{fig:3}
\end{figure}

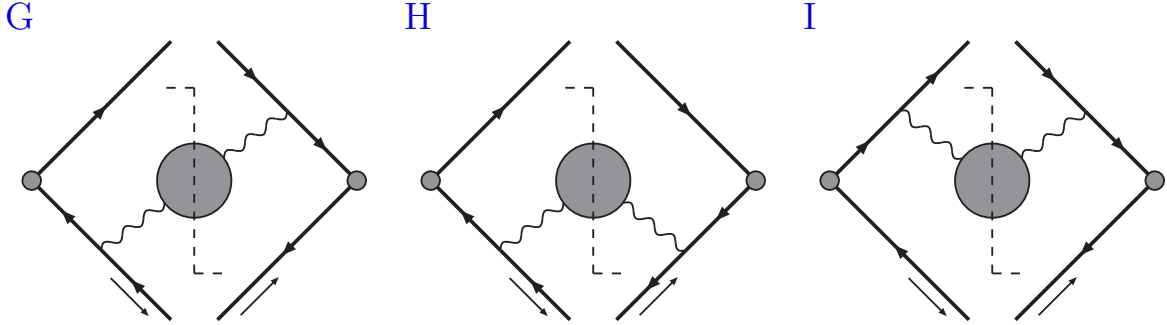
\begin{figure}[t]
\begin{center}
\fcolorbox{white}{white}{
\begin{picture}(200,150)(-20,0) % (-25,-30)
    \SetScale{0.35}

    \SetWidth{4}

    %fermion lines
    \ArrowLine(0,150)(150,300)

    \ArrowLine(75,75)(0,150)
    \ArrowLine(150,0)(75,75)

    \ArrowLine(350,150)(200,0)

    \ArrowLine(275,225)(350,150)
    \ArrowLine(200,300)(275,225)

    \SetWidth{2}

    %photon line
    \SetColor{Black}
    \COval(175,150)(40,40)(10){Black}{Gray}

    \Photon(207,174)(275,225){5}{3}
    \Photon(75,75)(143,126){5}{3}

    %cut lines
    \DashLine(175,250)(175,50){10}
    \DashLine(145,250)(175,250){10}
    \DashLine(205,50)(175,50){10}

    %Diagram label
    \Text(-10,110)[lb]{\large{\dg}}

    %momentum labels
    \LongArrow(85,45)(125,05)
    \LongArrow(225,05)(265,45)

    %vertex
    \SetColor{Black}
    \COval(350,150)(10,10)(10){Black}{Gray}
    \COval(0,150)(10,10)(10){Black}{Gray}
\end{picture}
\hspace{-2.0cm}
\begin{picture}(200,150)(-20,0) % (-25,-30)
    \SetScale{0.35}

    \SetWidth{4}

    %fermion lines

    \ArrowLine(0,150)(150,300)

    \ArrowLine(75,75)(0,150)
    \ArrowLine(150,0)(75,75)

    \ArrowLine(275,75)(200,0)
    \ArrowLine(350,150)(275,75)

    \ArrowLine(200,300)(350,150)

    \SetWidth{2}

    %photon line
    \SetColor{Black}
    \COval(175,150)(40,40)(10){Black}{Gray}

    \Photon(207,126)(275,75){5}{3}
    \Photon(75,75)(143,126){5}{3}

    %cut lines
    \DashLine(175,250)(175,50){10}
    \DashLine(145,250)(175,250){10}
    \DashLine(205,50)(175,50){10}

    %Diagram label
    \Text(-10,110)[lb]{\large{\dh}}

    %momentum labels
    \LongArrow(85,45)(125,05)
    \LongArrow(225,05)(265,45)

    %vertex
    \SetColor{Black}
    \COval(350,150)(10,10)(10){Black}{Gray}
    \COval(0,150)(10,10)(10){Black}{Gray}
\end{picture}
\hspace{-2.0cm}
\begin{picture}(200,150)(-20,0) % (-25,-30)
    \SetScale{0.35}

    \SetWidth{4}

    %fermion lines

    \ArrowLine(0,150)(75,225)
    \ArrowLine(75,225)(150,300)
    \ArrowLine(150,0)(0,150)

    \ArrowLine(350,150)(200,0)

    \ArrowLine(275,225)(350,150)
    \ArrowLine(200,300)(275,225)

    \SetWidth{2}

    %photon line
    \SetColor{Black}
    \COval(175,150)(40,40)(10){Black}{Gray}

    \Photon(75,225)(143,174){5}{3}
    \Photon(207,174)(275,225){5}{3}

    %cut lines
    \DashLine(175,250)(175,50){10}
    \DashLine(145,250)(175,250){10}
    \DashLine(205,50)(175,50){10}

    %Diagram label
    \Text(-10,110)[lb]{\large{\di}}

    %momentum labels
    \LongArrow(85,45)(125,05)
    \LongArrow(225,05)(265,45)

    %vertex
    \SetColor{Black}
    \COval(350,150)(10,10)(10){Black}{Gray}
    \COval(0,150)(10,10)(10){Black}{Gray}
\end{picture}
}
\end{center}
\caption{Diagrams \dg, \dh~and \di.  
These classes of diagrams contribute to integrals $I_{\dg}, I_{\dh}$, and $I_{\di}$ when the self-energy graphs involve gluons or ghosts and they contribute to integrals  $\tilde{I}_{\dg}, \tilde{I}_{\dh}$, and $\tilde{I}_{\di}$ when the self-energy graphs
involve a fermion/anti-fermion pair.}
\label{fig:4}
\end{figure}

\begin{figure}[h!]
\begin{center}
\fcolorbox{white}{white}{
\begin{picture}(200,150)(-20,0) % (-25,-30)
    \SetScale{0.35}

    \SetWidth{4}

    %fermion lines
    \ArrowLine(275,225)(350,150)
    \ArrowLine(200,300)(275,225)

    \ArrowLine(50,100)(0,150)
    \ArrowLine(100,50)(50,100)
    \ArrowLine(150,0)(100,50)

    \ArrowLine(0,150)(150,300)

    \ArrowLine(350,150)(200,0)

    \SetWidth{2}

    %photon line
    \Photon(50,100)(175,150){5}{4}
    \Photon(100,50)(175,150){5}{4}

    \Photon(175,150)(275,225){5}{4}

    %cut lines
    \DashLine(200,250)(200,80){10}
    \DashLine(170,250)(200,250){10}
    \DashLine(230,80)(200,80){10}

    %Diagram label
    \Text(-10,110)[lb]{\large{\dj}}

    %momentum labels
    \LongArrow(85,45)(125,05)
    \LongArrow(225,05)(265,45)

    \LongArrow(215,160)(255,190)
    \Text(85,50)[lb]{{\Black{$q$}}}

    \LongArrow(90,130)(130,148)
    \Text(30,55)[lb]{{\Black{$q-k$}}}

    \LongArrow(135,70)(170,110)
    \Text(57,25)[lb]{{\Black{$k$}}}

    %vertex
    \SetColor{Black}
    \COval(350,150)(10,10)(10){Black}{Gray}
    \COval(0,150)(10,10)(10){Black}{Gray}
\end{picture}
\hspace{-2cm}
\begin{picture}(200,150)(-20,0)
\SetScale{0.35}

    \SetWidth{4}

    %fermion lines
    \ArrowLine(0,150)(75,225)
    \ArrowLine(75,225)(150,300)

    \ArrowLine(75,75)(0,150)
    \ArrowLine(150,0)(75,75)

    \ArrowLine(350,150)(200,0)

    \ArrowLine(275,225)(350,150)
    \ArrowLine(200,300)(275,225)

    \SetWidth{2}

    %cut lines
    \DashLine(200,250)(200,80){10}
    \DashLine(170,250)(200,250){10}
    \DashLine(230,80)(200,80){10}

    %photon line
    \Photon(75,225)(175,150){5}{4}
    %\Photon(175,150)(275,75){5}{4}
    \Photon(175,150)(275,225){5}{4}
    \Photon(75,75)(175,150){5}{4}

    %Diagram label
    \Text(-10,110)[lb]{\large{\dk}}

    %momentum labels
        \LongArrow(85,45)(125,05)
        \LongArrow(225,05)(265,45)

    \LongArrow(110,80)(150,110)
        \Text(40,20)[lb]{{\Black{$q-k$}}}

    \LongArrow(110,220)(145,190)
      \Text(43,80)[lb]{{\Black{$k$}}}

    \LongArrow(215,160)(255,190)
        \Text(85,50)[lb]{{\Black{$q$}}}

    %vertex
        \SetColor{Black}
    \COval(350,150)(10,10)(10){Black}{Gray}
    \COval(0,150)(10,10)(10){Black}{Gray}
\end{picture}
\hspace{-2cm}
\begin{picture}(200,150)(-20,0) % (-25,-30)
    \SetScale{0.35}

    \SetWidth{4}

    %fermion lines
    \ArrowLine(0,150)(150,300)

    \ArrowLine(75,75)(0,150)
    \ArrowLine(150,0)(75,75)

    \ArrowLine(350,150)(200,0)

    \ArrowLine(275,225)(350,150)
    \ArrowLine(200,300)(275,225)

    \SetWidth{2}

    %photon line
    \SetColor{Black}
    \COval(125,113)(20,20)(10){Black}{White}
    \Line(145,113)(105,113)
    \Line(125,93)(125,133)
    
    \Photon(139,127)(275,225){5}{5.5}
    \Photon(75,75)(111,99){5}{1.5}

    %\Photon(125,113)(275,225){5}{6}
    %\Photon(75,75)(125,113){5}{2}

    %cut lines
    \DashLine(195,250)(195,80){10}
    \DashLine(195,250)(170,250){10}
    \DashLine(220,80)(195,80){10}

    %Diagram label
    \Text(-10,110)[lb]{\large{\dl}}

    %momentum labels
    \LongArrow(85,45)(125,05)
    \LongArrow(225,05)(265,45)

    %vertex
    \SetColor{Black}
    \COval(350,150)(10,10)(10){Black}{Gray}
    \COval(0,150)(10,10)(10){Black}{Gray}
\end{picture}
}
\end{center}
\caption{Diagrams \dj~and \dk~account for the interference between the one-loop virtual emission amplitude 
and the single gluon emission amplitude. Diagram \dl~is the contribution from charge renormalization.}
\label{fig:5}
\vspace{5cm}
\end{figure}
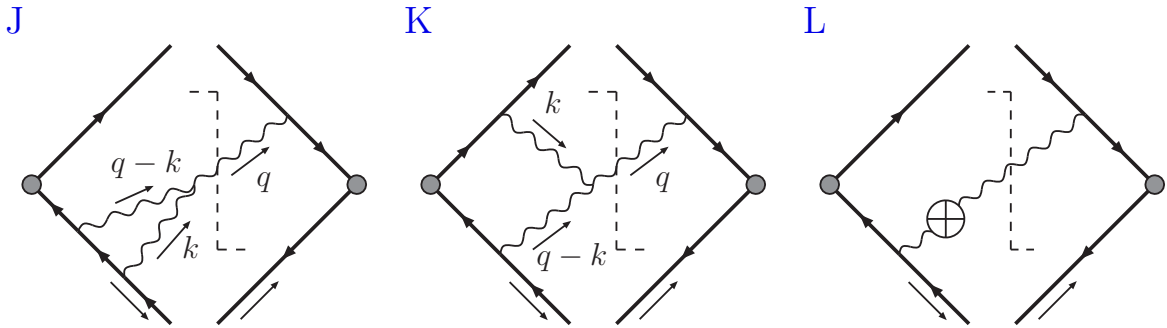

The order $\alpha_s^2$ calculation involves pure virtual graphs, pure real emission graphs, and 
interference between the two.  
The pure virtual contributions to the soft function give scaleless integrals which convert
IR divergences to UV divergences, and are not explicitly written.  The diagrams needed to compute the pure real emission
contributions are shown in Figs.~\ref{fig:1}-\ref{fig:4}, whereas the interference graphs between
the order $\alpha_s$ real and virtual emission amplitudes are shown in Fig.~\ref{fig:5}.  

The integrals corresponding to the two diagrams in Fig.~\ref{fig:1} 
(and the twin of diagram \da~obtained by interchanging $k$ and $q$) are given,  in $d = 4-2\e$ dimensions using
the $\overline{\text{MS}}$ scheme, by
%----------------------------------------------------------------------------------------------------
\begin{align}
&I_{\da} 
  = 2 
    (-4g^4) \left( C_F^2 -  \frac{C_F C_A}{2}\right) 
    \left(
        \frac{\mu^2 e^{\gamma_E}}{4\pi}
    \right)^{2\e}
    \int \frac{\rd^d q}{(2\pi)^d}
    \int \frac{\rd^d k}{(2\pi)^d}
    \frac{1}{k^- k^+ q^- q^+ }
    F(k_L, k_R)  
\end{align}
%----------------------------------------------------------------------------------------------------
and
%----------------------------------------------------------------------------------------------------
\begin{align}
&I_{\db} 
  = 
    -4g^4 
    \left(
        \frac{\mu^2 e^{\gamma_E}}{4\pi}
    \right)^{2\e}
    \int \frac{\rd^d q}{(2\pi)^d}
    \int \frac{\rd^d k}{(2\pi)^d}
    \frac{1}{(k^+ + q^+) (k^- + q^-)  }
\nn
& \qquad \times 
    \left\{ 
      \left( C_F^2 -  \frac{C_F C_A}{2}\right) 
      \left( \frac{1}{k^-k^+}  + \frac{1}{q^- q^+} \right)
      +
      C_F^2
      \left( \frac{1}{k^-q^+}  + \frac{1}{q^- k^+} \right)
      \right\}
    F(k_L, k_R) , 
\end{align}
%----------------------------------------------------------------------------------------------------
where $k^- = \nbar \cdot k$, $k^+ = n \cdot k$ and $F(k_L, k_R)$ contains the $\delta(q^2)$ and $\delta(k^2)$ factors which
put the emitted gluons on shell and the phase-space restrictions in the definition of the hemisphere soft function.  Explicitly, $F(k_L, k_R)$ is given by
%----------------------------------------------------------------------------------------------------
\begin{align}
\label{cut1}
F(k_L, k_R)  
&= \frac{1}{2!}
    (-2\pi i)^2 \delta(k^2) \delta(q^2)
\nn
&\times 
    \Bigl[
      \Theta( k^- - k^+ )
      \Theta( q^+ - q^- )
      \delta( k^+ - k_R)
      \delta( q^- - k_L)
\nn
&\ \   + 
      \Theta( k^+ - k^- ) 
      \Theta( q^- - q^+ ) 
      \delta( k^- - k_L) 
      \delta( q^+ - k_R) 
\nn
&\ \   + 
      \Theta( k^- - k^+ ) 
      \Theta( q^- - q^+ ) 
      \delta( k^+ + q^+ - k_R)
      \delta(k_L)
\nn
&\ \   + 
      \Theta( q^+ - q^- ) 
      \Theta( k^+ - k^- ) 
      \delta( k^- + q^- - k_L)
      \delta(k_R)
    \Bigr] .
\end{align}
%----------------------------------------------------------------------------------------------------
In each diagram, the momentum has been routed so that the 4-vectors $k$ and $q$ correspond to the 
final state gluons. The gluonic contribution to the $C_F C_A$ color factor will be symmetric in 
$k$ and $q$ due to the fact that the radiated gluons are identical particles. In the first
diagram, a factor of two has been added since the integrand is unchanged after $k \leftrightarrow q$,
whereas in the second diagram, the SCET Feynman rules for two gluon emission from a single soft Wilson line
automatically account for $k\leftrightarrow q$. The factor of $1/2!$ needed for averaging over 
$k \leftrightarrow q$ is in $F(k_L, k_R)$.  Not shown in Fig.~\ref{fig:1} is the graph 
that corresponds to the complex conjugate of diagram \db.  This
diagram gives the same integral as diagram \db. Since we are interested in the $C_F C_A$ contribution, 
the linear combination of interest is $I_{\da} + 2I_{\db}$.  Diagram \da~and its identical twin are 
self-conjugate and only contribute once because they represent the squares of tree-level Feynman diagrams.

%-----------------------------------------------------------------

%------------FIGURE------------------------------------------------
%-----------------------------------------------------------------
%------------FIGURE------------------------------------------------
%-----------------------------------------------------------------

%---- next page

%------------FIGURE------------------------------------------------
%-----------------------------------------------------------------

%------------FIGURE------------------------------------------------
%-----------------------------------------------------------------

There are four classes of diagrams involving the triple gauge coupling.  Diagrams \dc~and \dd, shown in Fig.~\ref{fig:2},
give the following integrals, 
%----------------------------------------------------------------------------------------------------
\begin{align}
I_{\dc} &= 
    -g^4 C_A C_F
    \left(
        \frac{\mu^2 e^{\gamma_E}}{4\pi}
    \right)^{2\e}
    \int \frac{\rd^d q}{(2\pi)^d}
    \int \frac{\rd^d k}{(2\pi)^d}
    \frac{F(k_L, k_R)}{(k^- + q^-)(k+q)^2 }
    \left(
    \frac{k^- + 2q^-}{ k^+ q^- }
    +
    \frac{q^- + 2k^-}{ q^+ k^- }
    \right)
\end{align}
%----------------------------------------------------------------------------------------------------
%----------------------------------------------------------------------------------------------------
\begin{align}
I_{\dd} 
  &= 
    -g^4 C_A C_F
    \left(
        \frac{\mu^2 e^{\gamma_E}}{4\pi}
    \right)^{2\e}
    \int \frac{\rd^d q}{(2\pi)^d}
    \int \frac{\rd^d k}{(2\pi)^d}
    \frac{F(k_L, k_R)}{(k^+ + q^+)  (k+q)^2 }
    \left(
    \frac{2k^+ + q^+}{k^+ q^- }
    +
    \frac{2q^+ + k^+}{q^+ k^- }
    \right)
\end{align}
%----------------------------------------------------------------------------------------------------
whereas, diagrams \de~and \df, shown in Fig~\ref{fig:3}, give 
%----------------------------------------------------------------------------------------------------
\begin{multline}
I_{\de} 
  = 
    g^4 
    C_F C_A
    \left(
        \frac{\mu^2 e^{\gamma_E}}{4\pi}
    \right)^{2\e}
    \int \frac{\rd^d q}{(2\pi)^d}
    \int \frac{\rd^d k}{(2\pi)^d}
\nn
 \qquad \times
    \left( \frac{1}{q^-}  - \frac{1}{k^-} \right)
    \frac{q^- - k^- }{(k^+ + q^+) (k^- + q^-) (k+q)^2  }
    F(k_L, k_R)
\end{multline}
%----------------------------------------------------------------------------------------------------
%----------------------------------------------------------------------------------------------------
\begin{multline}
I_{\df} 
 = 
    g^4 
    C_F C_A
    \left(
        \frac{\mu^2 e^{\gamma_E}}{4\pi}
    \right)^{2\e}
    \int \frac{\rd^d q}{(2\pi)^d}
    \int \frac{\rd^d k}{(2\pi)^d}
\nn
 \qquad \times
    \left( \frac{1}{k^+}  - \frac{1}{q^+} \right)
    \frac{k^+ - q^+ }{(k^+ + q^+) (k^- + q^-) (k+q)^2  }
    F(k_L, k_R)
\end{multline}
%----------------------------------------------------------------------------------------------------
Each of these diagrams has a complex conjugate and so they contribute twice.

There are three self-energy topologies, shown in Fig.~\ref{fig:4}.  The gluon and ghost self-energy graphs contribute to integrals $I_{\dg}, I_{\dh}$ and $I_{\di}$ below.  
%----------------------------------------------------------------------------------------------------
\begin{multline}
I_{\dg} 
= 
    g^4 C_F C_A
    \left(
        \frac{\mu^2 e^{\gamma_E}}{4\pi}
    \right)^{2\e}
    \int \frac{\rd^d q}{(2\pi)^d}
    \int \frac{\rd^d k}{(2\pi)^d}
    \frac{1}{(k^-+q^-)(k^++q^+)(k+q)^4 }
 \nn
  \qquad \times 
    \Bigl[ 
         q^+ [(d-6)q^- - (d+2)k^- ] 
        +k^+ [(d-6)k^- - (d+2)q^- ] 
        +16 k\cdot q 
    \Bigr]
    F(k_L, k_R)
\end{multline}
%----------------------------------------------------------------------------------------------------
%----------------------------------------------------------------------------------------------------
\begin{multline}
I_{\dh}
  = 
    g^4 C_F C_A
    \left(
        \frac{\mu^2 e^{\gamma_E}}{4\pi}
    \right)^{2\e}
    \int \frac{\rd^d q}{(2\pi)^d}
    \int \frac{\rd^d k}{(2\pi)^d}
    \frac{1}{(k^-+q^-)^2(k+q)^4 }
 \nn
 \qquad \times 
    \Bigl[ 
         2(d+2) q^-k^-  
         -(d-6) (k^-)^2
         -(d-6) (q^-)^2
    \Bigr]
    F(k_L, k_R)
\end{multline}
%----------------------------------------------------------------------------------------------------
%----------------------------------------------------------------------------------------------------
\begin{multline}
I_{\di} 
  = 
    g^4 C_F C_A
    \left(
        \frac{\mu^2 e^{\gamma_E}}{4\pi}
    \right)^{2\e}
    \int \frac{\rd^d q}{(2\pi)^d}
    \int \frac{\rd^d k}{(2\pi)^d}
    \frac{1}{(k^+ + q^+)^2(k+q)^4 }
 \nn
 \qquad \times 
    \Bigl[ 
         2(d+2) q^+k^+  
         -(d-6) (k^+)^2
         -(d-6) (q^+)^2
    \Bigr]
    F(k_L, k_R)
\end{multline}
%----------------------------------------------------------------------------------------------------
As usual, cutting Feynman diagrams removes any symmetry factors that were associated to the cut lines prior
to cutting.  It is also worth reminding the reader that, in order to consistently combine the ghost emission
diagrams with the gluon emission diagrams, we have to double-count the ghosts 
(they do not have the $1/2!$ symmetry factor that the gluons do). Diagram \dg~has a complex conjugate 
graph which must be included but diagrams \dh~and \di, like diagram \da, represent squares of tree-level
Feynman diagrams and are therefore self-conjugate.

Adding all of these contributions together, we have
%----------------------------------------------------------------------------------------------------
\begin{align}
\label{int:CA}
&S^{R}_{C_A}(k_L, k_R)
= I_{\da} + I_{\dh} + I_{\di} + 2(I_{\db}+I_{\dc}+I_{\dd}+I_{\de}+I_{\df}+I_{\dg}) \qquad (\text{$C_F C_A$ part})
\nn
&\qquad
    =   g^4 C_F C_A
    \left(
        \frac{\mu^2 e^{\gamma_E}}{4\pi}
    \right)^{2\e}
    \int \frac{\rd^d q}{(2\pi)^d}
    \int \frac{\rd^d k}{(2\pi)^d}
    \Bigl\{ 
    \frac{2}
    {
      (k\cdot q)^2 
      k^{-} k^{+} q^{-} q^{+} 
      (k^{-}+q^{-}) (k^{+}+q^{+})
    }
\nn
&\qquad \times 
   \Bigl[ 
    -k \cdot q 
    \Bigl(
         (k^{-})^2 q^{+} ( 2 k^{+}+q^{+} )
        +2 k^{-} q^{-} 
         \left(
            (k^{+})^2 - k^{+} q^{+}+(q^{+})^2
         \right)
        +k^{+}(q^{-})^2 (k^{+}+2 q^{+})
    \Bigr)
\nn
&\qquad
    +2 (k \cdot q)^2 
        \Bigl( k^{-} (2 k^{+}+q^{+})
         +q^{-} (k^{+}+2q^{+})
        \Bigr)
  \Bigr] 
\nn
&\qquad
  +
  (\epsilon -1)
    \frac{2(k^{+} q^{-}-k^{-} q^{+})^2}
    {
      (k\cdot q)^2 
      (k^{-}+q^{-})^2 (k^{+}+q^{+})^2
    }
    \Bigr\} 
  F(k_L, k_R) \,.
\end{align}
%----------------------------------------------------------------------------------------------------

Before presenting the result for the $C_F C_A$ color factor, we briefly describe our general computational strategy. Normally, one expects scaleless integrals to be simpler than single scale integrals. In this particular case, the single scale integrals (with scale $k_L/k_R$) are actually much less technically demanding. This is true primarily because these contributions (see the first two terms of Eq.~(\ref{cut1})) are integrable at $\epsilon = 0$. It turns out that this special feature of the problem more than makes up for the fact that single scale integrals are generically harder to evaluate than scaleless integrals.

The calculation proceeds as follows for a single scale integral. First there will be an integral over angles (the integrand depends non-trivially on $k \cdot q$) that can be done analytically to all orders in $\epsilon$. It is then convenient to Taylor series expand the resulting hypergeometric functions using the HypExp package~\cite{Huber:2005yg} for {\sc Mathematica}. In fact, the whole integrand can be expanded in a Taylor series in $\epsilon$ and integrated term-by-term, due to the fact that the integral converges at $\epsilon = 0$. With a modest amount of knowledge of the basic functional identities satisfied by the polylogarithm functions, it is possible to do the resulting two-fold one-parameter integral in {\sc Mathematica} and express the final result in terms of a minimal basis of transcendental functions. The results of our single scale calculations for both non-trivial color factors are tabulated in the Appendix.

The evaluation of a scaleless integral (originating from the last two terms of Eq.~(\ref{cut1})) begins in much the same way. Unfortunately, it quickly becomes clear that what remains after integrating over all angles has a non-trivial analytical structure (considered as a function of $\epsilon$). In particular, the integral diverges at $\epsilon = 0$. Expanding an integral of this class under the integral sign is significantly more complicated and requires new tools. To begin, one should transform all hypergeometric functions in the integrand and expose their singularity structure. In this fashion, one learns that there is a line of singularities within the region of integration. A well-known procedure called sector decomposition~\cite{Heinrich:2008si} allows one to move singularities within the region of integration to singularities on the boundaries of the region of integration. Sector decomposition works as follows. Through a sequence of variable changes and interchanges of integration orders, all phase-space singularities are put into a canonical form. At this point, one can use an expansion in distributions to extract singularities in $\epsilon$ under the integral sign. Finally, the entire integrand can be expanded in $\epsilon$ in terms of distributions and ordinary functions and one can integrate the Laurent series term-by-term.

%Sector decomposition is well-suited to deal with the class of integrals that arise from the last two terms of Eq.~(\ref{cut1}). In fact, after remapping the line of singularities as described above, one of the three remaining integrations factorizes and a nice reduction of the problem occurs. At this point, we are faced with a number of sub-problems. That is to say, what remains after integrating out the variable that factorized are some number of simple two-fold integrals over various combinations of delta functions and plus distributions in the integrand. Each of these sub-problems can be dealt with by evaluating one of the integrals analytically and then using {\sc Mathematica} to evaluate the final integral numerically to high precision. At each order in $\epsilon$, all contributions were summed and then the LLL algorithm (if unfamiliar check the documentation for the {\sc Mathematica} function {\sc LatticeReduce}) was employed to recover a relation between the numerical results obtained and a simple basis of transcendental numbers: \{$1,\pi^2, \zeta_3, \pi^4$\}.

Once we understood the computational procedure described above, it was straightforward to evaluate the integrals of interest for the $C_F C_A$ color factor. The result has the form
%----------------------------------------------------------------------------------------------------
\begin{equation}
\label{result:CA}
S^{R}_{C_A}(k_L, k_R) 
=  
    \left(\frac{\alpha}{4\pi} \right)^2 
    C_F C_A
    \left[
      \frac{\mu^{4\e}}{( k_R k_L)^{1+2\e}} 
      f_{C_A}\left(\frac{k_L}{k_R},\e\right) 
    +
    \left(
      \frac{\mu^{4\e}}{k_L^{1+4\e}}  \delta(k_R)
      +
      \frac{\mu^{4\e}}{k_R^{1+4\e}}  \delta(k_L)
      \right) g_{C_A}(\e)
    \right]. \non
\end{equation}
%----------------------------------------------------------------------------------------------------
The first term corresponds to the first two terms in Eq.~(\ref{cut1}), those that account for the possibility that exactly one gluon is radiated into each hemisphere. It depends on $f_{C_A}(r,\e)$, a dimensionless function of $r = k_L/k_R$ and $\e$. It can be written as an expansion
in $\e$ as
%----------------------------------------------------------------------------------------------------
\begin{align}
\label{fca:exp}
f_{C_A}(r,\e) 
&= 
       f_{C_A}^{(0)}(r)
+ \e   f_{C_A}^{(1)}(r)
+ \e^2 f_{C_A}^{(2)}(r) + \cdots .
\end{align}
%----------------------------------------------------------------------------------------------------
The expressions for $f_{C_A}^{(n)}(r)$ are quite lengthy and are given in the appendix for $n = 0, 1, 2$.
The second term in Eq.~(\ref{result:CA}) accounts for the fact that both gluons can propagate into
the same hemisphere and it has no non-trivial $k_L$ or $k_R$ dependence. $g_{C_A}(\e)$ is simply a constant with $\epsilon$ expansion  
%----------------------------------------------------------------------------------------------------
\begin{align}
g_{C_A}(\e)  \label{gca}
&=  
\frac{4}{\epsilon^3}
+\frac{22}{3 \epsilon ^2}
+\frac{1}{\epsilon} \left( \frac{134}{9} -\frac{4 \pi ^2}{3} \right)
-\frac{116 \zeta_3}{3}
+\frac{11 \pi ^2}{9}
+\frac{772}{27}
\nn
& \qquad
+\left(\frac{484 \zeta_3}{9}
+\frac{4784}{81}
+\frac{67 \pi ^2}{27}
-\frac{137 \pi^4}{90}\right) \epsilon . 
\end{align}
%----------------------------------------------------------------------------------------------------

The interference between the one-loop and tree-level single gluon emission amplitudes is
shown in diagrams \dj~and \dk~of Fig.~\ref{fig:5}. 
%Diagram \dl~incontributes to both the $C_F C_A$ and $C_F n_f T_F$ color factors.
The integrals associated with diagram \dj~are scaleless and are set to zero in dimensional
regularization.  Diagram \dk~gives the integral
%----------------------------------------------------------------------------------------------------
\begin{align}
I_{\dV}
&= 
  4 
  (-g^4) C_A C_F
  \int \frac{\rd^d q}{(2\pi)^d}
  \frac{1}{q^- }
  \int \frac{\rd^d k}{(2\pi)^d}
  \frac{2q^- - k^-}{k^+  (q^- - k^-)  (q-k)^2 k^2 }
\nn
&\qquad \times
    (-2\pi i) \delta(q^2) 
    [ \Theta( q^- - q^+ ) \delta( q^+ - k_R ) \delta( k_L ) 
      +
      \Theta( q^+ - q^- ) \delta( q^- - k_L ) \delta( k_R )  ].
\end{align}
%----------------------------------------------------------------------------------------------------
There are 2 diagrams with the topology of diagram \dk. When they are considered with single real emission
phase-space cuts, they can easily be mapped into each other and therefore give identical results.
Both diagrams also have a complex conjugate graph and these obviously give equal contributions as well.
This is why $I_{\dV}$ has an overall factor of 4 out front.  
After evaluating this integral, the real-virtual interference contribution becomes
%----------------------------------------------------------------------------------------------------
\begin{align}
\label{result:virt}
S^{\rm V}_{C_A}(k_L, k_R) 
&=  
    \left(\frac{\alpha}{4\pi} \right)^2 
    C_F C_A
    \left(
      \frac{\mu^{4\e}}{k_L^{1+4\e}}  \delta(k_R)
      +
      \frac{\mu^{4\e}}{k_R^{1+4\e}}  \delta(k_L)
      \right) v_{C_A}(\e) ,
\end{align}
%----------------------------------------------------------------------------------------------------
where $v_{C_A}(\e)$ can be expanded in $\e$ as
%----------------------------------------------------------------------------------------------------
\begin{align} \label{vca}
v_{C_A}(\e)
&=  
-\frac{4}{\epsilon^3}
+\frac{2\pi^2}{\epsilon} 
+\frac{32 \zeta_3}{3}
-\e \frac{\pi ^4}{30} .
\end{align}
%----------------------------------------------------------------------------------------------------
It is worth noting that, in this case, the application of the optical theorem for Feynman diagrams is a bit subtle;
one finds an explicit factor of ${\rm exp}(\pm i \pi \epsilon)$ after doing the $k$ integral
(the sign of the phase depends on the precise pole prescription). 
Cutkosky's rules still apply provided that one keeps only the appropriate projection of the complex phase.
After a moment's thought it becomes clear that the real part, $\cos(\pi \epsilon)$
({\it independent} of the pole prescription), is what one needs to keep to complete
the calculation and derive the above result. 

The result of diagram \dl, including the complex conjugate
graph, is given by 
%----------------------------------------------------------------------------------------------------
\begin{align}
\label{result:ct}
S^{\rm Ren }(k_L, k_R) 
&=  
    -
    \left(\frac{\alpha}{4\pi} \right)^2 
    C_F 
    \left(
      \frac{\mu^{2\e}}{k_L^{1+2\e}}  \delta(k_R)
      +
      \frac{\mu^{2\e}}{k_R^{1+2\e}}  \delta(k_L)
      \right) 
      \frac{4e^{\gamma_E}}{\e^2\Gamma(1-\e)}
      \beta_0
\end{align}
%----------------------------------------------------------------------------------------------------
where $\beta_0 =  \frac{11}{3} C_A - \frac{4}{3} n_f T_F$ is the first expansion coefficient of the 
QCD $\beta$-function, $\beta(g)/g = \frac{\alpha_s}{4\pi} \beta_0$.  Finally, the total contribution to the
$C_F C_A$ color factor is given by
%----------------------------------------------------------------------------------------------------
\begin{align}
\label{CAtotal}
S_{C_A}(k_L, k_R) 
  &=  
    S^{R}_{C_A}(k_L, k_R) 
    +
    S^{V}_{C_A}(k_L, k_R) 
    +
    S^{\rm Ren}_{C_A}(k_L, k_R), 
\end{align}
%----------------------------------------------------------------------------------------------------
where $S^{\rm Ren}_{C_A}$ is the $C_F C_A$ part of $S^{\rm Ren}$.

\subsection{$C_F n_F T_F$ color structure}
The diagrams involving a fermion loop contribute to the $C_F n_f T_F$ color factor and give 
integrals $\tilde{I}_{\dg}, \tilde{I}_{\dh},$ and $\tilde{I}_{\di}$.  
The first topology in Fig.~\ref{fig:4}, where the blob now represents a fermion loop, gives 
%----------------------------------------------------------------------------------------------------
\begin{align}
\tilde{I}_{\dg} 
  &= 
    g^4 C_F n_f T_F
    \left(
        \frac{\mu^2 e^{\gamma_E}}{4\pi}
    \right)^{2\e}
    \int \frac{\rd^d q}{(2\pi)^d}
    \int \frac{\rd^d k}{(2\pi)^d}
    \frac{4(k^+ q^- + k^- q^+ - 2k \cdot q)}{(k^+ + q^+) (k^- + q^-) (k+q)^4  }
 F_{n_f}( k_L, k_R) .
\end{align}
%----------------------------------------------------------------------------------------------------
The phase-space cut is accounted for by 
%----------------------------------------------------------------------------------------------------
\begin{multline}
\label{cut2}
F_{n_f}(k_L, k_R)  
= 
    (-2\pi i)^2 \delta(k^2) \delta(q^2)
\nn
\times 
    \Bigl[
      \Theta( k^- - k^+ )
      \Theta( q^+ - q^- )
      \delta( k^+ - k_R)
      \delta( q^- - k_L)
%\nn
%&\ \   
+ 
      \Theta( k^+ - k^- ) 
      \Theta( q^- - q^+ ) 
      \delta( k^- - k_L) 
      \delta( q^+ - k_R) 
\nn
\ \   + 
      \Theta( k^- - k^+ ) 
      \Theta( q^- - q^+ ) 
      \delta( k^+ + q^+ - k_R)
      \delta(k_L)
%\nn
%&\ \ 
  + 
      \Theta( q^+ - q^- ) 
      \Theta( k^+ - k^- ) 
      \delta( k^- + q^- - k_L)
      \delta(k_R)
    \Bigr].
\end{multline}
%----------------------------------------------------------------------------------------------------
The complex conjugate of this diagram gives the same result, so $\tilde{I}_{g}$ contributes twice.
For the second and third topologies shown in Fig.~\ref{fig:4}, we get
%----------------------------------------------------------------------------------------------------
\begin{equation}
\tilde{I}_{\dh} 
  =
    g^4 C_F n_f T_F
    \left(
        \frac{\mu^2 e^{\gamma_E}}{4\pi}
    \right)^{2\e}
    \int \frac{\rd^d q}{(2\pi)^d}
    \int \frac{\rd^d k}{(2\pi)^d}
    \frac{-8 k^+ q^+}{(k^+ + q^+)^2 (k+q)^4  }
 F_{n_f}( k_L, k_R) 
\end{equation}
%----------------------------------------------------------------------------------------------------
and
%----------------------------------------------------------------------------------------------------
\begin{equation}
\tilde{I}_{\di} 
  =
    g^4 C_F n_f T_F
    \left(
        \frac{\mu^2 e^{\gamma_E}}{4\pi}
    \right)^{2\e}
    \int \frac{\rd^d q}{(2\pi)^d}
    \int \frac{\rd^d k}{(2\pi)^d}
    \frac{-8 k^- q^-}{(k^- + q^-)^2 (k+q)^4  }
 F_{n_f}( k_L, k_R) \,.
\end{equation}
%----------------------------------------------------------------------------------------------------
The sum of these contributions is
%----------------------------------------------------------------------------------------------------
\begin{align}
\label{int:nf}
&S^{R}_{n_f}(k_L, k_R)
= 2\tilde{I}_{\dg}  + \tilde{I}_{\dh} + \tilde{I}_{\di} 
\nn
&\qquad
 =
 g^4 C_F n_f T_F
    \left(
        \frac{\mu^2 e^{\gamma_E}}{4\pi}
    \right)^{2\e}
 \int \frac{\rd^d q}{(2\pi)^d}
 \int \frac{\rd^d k}{(2\pi)^d}
\nn
&\qquad \times
 \frac{8}{(k+q)^4  }
 \left( 
     \frac{k^+ q^- + k^- q^+ - 2k \cdot q}{(k^+ + q^+) (k^- + q^-)}
     -
     \frac{k^- q^- }{(k^- + q^-)^2}
     -
     \frac{k^+ q^+ }{(k^+ + q^+)^2}
 \right)
 F_{n_f}( k_L, k_R) 
\end{align}
%----------------------------------------------------------------------------------------------------
Evaluating this integral gives
%----------------------------------------------------------------------------------------------------
\begin{equation}
\label{result:nf}
S^{R}_{n_f}(k_L, k_R) 
=  
    \left(\frac{\alpha}{4\pi} \right)^2 
    C_F n_f T_F
    \left[
      \frac{\mu^{4\e}}{( k_R k_L)^{1+2\e}} 
      f_{n_f}\left(\frac{ k_L}{ k_R},\e\right ) 
    +
    \left(
      \frac{\mu^{4\e}}{k_L^{1+4\e}}  \delta(k_R)
      +
      \frac{\mu^{4\e}}{k_R^{1+4\e}}  \delta(k_L)
      \right) g_{n_f}(\e)
    \right].  \non
\end{equation}
%----------------------------------------------------------------------------------------------------
As in the $C_F C_A$ case, the first term corresponds to the quark and anti-quark propagating into different 
hemispheres and it depends on $r = k_L/k_R$ in a non-trivial way through a function $f_{n_f}(r,\e)$. $f_{n_f}(r,\e)$ can be expanded in a Taylor series in $\e$ as
%----------------------------------------------------------------------------------------------------
\begin{align}
\label{fnf:exp}
f_{n_f}(r,\e) 
&= 
       f_{n_f}^{(0)}(r)
+ \e   f_{n_f}^{(1)}(r) + \cdots .
\end{align}
%----------------------------------------------------------------------------------------------------
The expressions for $f_{n_f}^{(n)}(r)$ are given in the appendix for $n= 0, 1$. For the $C_F n_f T_F$ color factor $n = 2$ plays no role due to the fact that $f_{n_f}^{(n)}(0) = f_{n_f}^{(n)}(\infty) = 0$.

The second term Eq.~(\ref{result:nf}) is present because both the quark and anti-quark may propagate into the same hemisphere as well. As before, this contribution has no non-trivial $k_L$ or $k_R$ dependence.  The constant $g_{n_f}$ has a series expansion
%----------------------------------------------------------------------------------------------------
\begin{equation} \label{gnf}
g_{n_f} (\e)
=  
-\frac{8}{3\epsilon^2}
-\frac{40}{9\epsilon}
-\frac{152}{27}
-\frac{4\pi ^2}{9}
+\left(
  -\frac{952}{81}
  -\frac{20 \pi ^2}{27}
  -\frac{176 \zeta_3}{9}
\right) \epsilon . 
\end{equation}
%----------------------------------------------------------------------------------------------------
The final contribution to this color factor is from the charge renormalization, diagram \dl, the 
results of which were given in Eq.~(\ref{result:ct}). Adding this contribution
to the real emission contributions yields the final result for the $C_F n_f T_F$ color factor.
It is 
%----------------------------------------------------------------------------------------------------
\begin{align}
\label{CFtotal}
S_{n_f}(k_L, k_R) 
  &=  
    S^{R}_{n_f}(k_L, k_R) 
    +
    S^{\rm Ren}_{n_f}(k_L, k_R),
\end{align}
%----------------------------------------------------------------------------------------------------
where $S^{\rm Ren}_{n_f}$ is the $C_F n_f T_F$ part of $S^{\rm Ren}$.

%%%%%%%%%%%%%%%%%%%%%%%%%%%%%%%%%%%%%%%%%%%%%%%%%%%%%%%%%%%%%%%%%%%%%%%%%%%%%%%%%
%%%%%%%%%%%%%%%%%%%%%%%%%%%%%%%%%%%%%%%%%%%%%%%%%%%%%%%%%%%%%%%%%%%%%%%%%%%%%%%%%
%%%%%%%%%%%%%%%%%%%%%%%%%%%%%%%%%%%%%%%%%%%%%%%%%%%%%%%%%%%%%%%%%%%%%%%%%%%%%%%%%
%%%%%%%%%%%%%%%%%%%%%%%%%%%%%%%%%%%%%%%%%%%%%%%%%%%%%%%%%%%%%%%%%%%%%%%%%%%%%%%%%

\subsection{Summary of the Calculation \label{sec:summary}}
In summary, we found that the 2-loop hemisphere soft function in $d=4-2\epsilon$ dimensions
has the form
%----------------------------------------------------------------------------------------------------
\begin{align}
\label{result:summ}
S(k_L, k_R,\mu) 
&=  
    \left(\frac{\alpha}{4\pi} \right)^2 
    \left[
      \frac{\mu^{4\e}}{( k_R k_L)^{1+2\e}} f\left(\frac{k_L}{k_R},\e\right) 
    +
    \left(
      \frac{\mu^{4\e}}{k_L^{1+4\e}}  \delta(k_R)
      +
      \frac{\mu^{4\e}}{k_R^{1+4\e}}  \delta(k_L)
      \right) h(\e)
    \right. \non \\
&    
\left.
    - 4C_F\beta_0 \left(
      \frac{\mu^{2\e}}{k_L^{1+2\e}}  \delta(k_R)
      +
      \frac{\mu^{2\e}}{k_R^{1+2\e}}  \delta(k_L)
      \right)\frac{e^{\gamma_E}}{\e^2 \Gamma(1-\e)}
    \right]\,.
\end{align}
%----------------------------------------------------------------------------------------------------
Here $f(r,\e)=f(1/r,\e)$ is the opposite-direction contribution (where the two gluons or two quarks go into opposite
hemispheres) and $h(\e)$ is the same-direction contribution.  Since all the $\mu$ dependence is shown explicitly,
$h(\e)$ cannot depend on $k_L$ or $k_R$ by dimensional analysis. The second line is the contribution that comes from the interference of the first non-trivial term in the expansion of the charge renormalization constant and the $\mathcal{O}(\alpha_s)$ hemisphere soft function. It is proportional to $\beta_0 = \frac{11}{3} C_A - \frac{4}{3} T_F n_f$.

There are 3 color structures, $C_F^2, C_F C_A$ and $C_F n_f T_F$. The $C_F^2$ color structure is trivial 
-- by non-Abelian exponentiation it is the square of the one-loop result. 
For the other two color structures the function $f(r,\e)$ is complicated. In both cases it is finite at $\e=0$, and in
the $C_F n_f T_F$ case, $f_{n_f}(0,\e) = f_{n_f}(\infty,\e) = 0$. We write
%----------------------------------------------------------------------------------------------------
\begin{align}
f(r,\e) &= f^{(0)}(r) + \e f^{(1)}(r) + \e^2 f^{(2)}(r) \label{fexp}
\end{align}
%----------------------------------------------------------------------------------------------------
The expansions in $\e$ of $f(r,\e)$ for the two color structures are given in 
the Appendix. Due to the fact that $f_{n_f}(0,\e) = f_{n_f}(\infty,\e) = 0$, $f^{(2)}_{n_f}(r)$ does not contribute
to the renormalized soft function and is not given.

For the same direction contribution, $h(\e)$, there are contributions from the real-emission diagrams
and, for the $C_F C_A$ color structure, interference between tree-level real emission and one-loop real-virtual graphs. 
The real emission contributions we called $g(\e)$, and are given in Eqs.~\eqref{gca} and~\eqref{gnf}.
The interference graphs are given by $v_{C_A}(\e)$ in Eq.~\eqref{vca}. Adding these terms we get
for the $C_F C_A$ color structure
%----------------------------------------------------------------------------------------------------
\begin{multline}
h_{C_A}(\e) = 
\frac{22}{3 \e^2}
+\frac{\frac{134}{9}+\frac{2 \pi ^2}{3}}{\e}
- 28 \zeta_3+\frac{11 \pi ^2}{9}+\frac{772}{27}
+\left(\frac{484 \zeta_3}{9}+\frac{4784}{81}+\frac{67 \pi ^2}{27} -\frac{14 \pi ^4}{9}\right) \e 
\end{multline}
%----------------------------------------------------------------------------------------------------
and for completeness, copying Eq.~\eqref{gnf}
%----------------------------------------------------------------------------------------------------
\begin{equation} 
h_{n_f}(\e)
=  
-\frac{8}{3\e^2}
-\frac{40}{9\e}
-\frac{152}{27}
-\frac{4\pi ^2}{9}
+\left(
  -\frac{952}{81}
  -\frac{20 \pi ^2}{27}
  -\frac{176 \zeta_3}{9}
\right) \e . 
\end{equation}
%----------------------------------------------------------------------------------------------------

%%%%%%%%%%%%%%%%%%%%%%%%%%%%%%%%%%%%%%%%%%%%%%%%%%%%%%%%%%%%%%%%%%%%%%%%%%%%%%%%%
%%%%%%%%%%%%%%%%%%%%%%%%%%%%%%%%%%%%%%%%%%%%%%%%%%%%%%%%%%%%%%%%%%%%%%%%%%%%%%%%%
%%%%%%%%%%%%%%%%%%%%%%%%%%%%%%%%%%%%%%%%%%%%%%%%%%%%%%%%%%%%%%%%%%%%%%%%%%%%%%%%%
%%%%%%%%%%%%%%%%%%%%%%%%%%%%%%%%%%%%%%%%%%%%%%%%%%%%%%%%%%%%%%%%%%%%%%%%%%%%%%%%%

%%%%%%%%%%%%%%%%%%%%%%%%%%%%%%%%%%%%%%%%%%%%%%%%%%%%%%%%%%%%%%%%%%%%%%%%%%%%%%%%%
%%%%%%%%%%%%%%%%%%%%%%%%%%%%%%%%%%%%%%%%%%%%%%%%%%%%%%%%%%%%%%%%%%%%%%%%%%%%%%%%%
%%%%%%%%%%%%%%%%%%%%%%%%%%%%%%%%%%%%%%%%%%%%%%%%%%%%%%%%%%%%%%%%%%%%%%%%%%%%%%%%%
%%%%%%%%%%%%%%%%%%%%%%%%%%%%%%%%%%%%%%%%%%%%%%%%%%%%%%%%%%%%%%%%%%%%%%%%%%%%%%%%%

\section{Integrating the soft function \label{sec:integrating}}
Now we would like to expand and renormalize the soft function. 
At one-loop, all that is necessary for the expansion is the relation
%----------------------------------------------------------------------------------------------------
\begin{align} \label{stardist}
\frac{\mu^{4\e} }{k^{1+2\e}} = 
  -\frac{1}{2\e} \delta(k) 
  + \left[ \frac{1}{k}  \right]_{\ast}
  - 2\e \left[ \frac{\ln \frac{k}{\mu} }{k}  \right]_{\ast} + \cdots,
\end{align}
%----------------------------------------------------------------------------------------------------
where the $\ast$-distributions are defined, for example, in \cite{Schwartz:2007ib}. Unfortunately,
this expansion cannot be used separately for $k_L$ and $k_R$, since the region where they both go to
zero is not well-defined. For example, what does $\delta(k_L)\delta(k_R)f(k_L/k_R)$ mean?
If we take $k_L \to 0$ first, then $k_R \to 0$, 
then we pick up $f(0)$. If we take $k_L,k_R \to 0$ holding $k_L = k_R$, then we pick up $f(1)$.
Unless $f(r)$ is constant, one must do the expansion more carefully.

A simple solution is just to expand in distributions of $p=k_L k_R$ and $r=k_L/k_R$. This expansion is well-defined,
and can be used to integrate any observable, such as thrust or heavy jet mass against the hemisphere soft function.
For example, consider the integrated soft function:
%----------------------------------------------------------------------------------------------------
\begin{align}
\cR(X,Y,\mu) \equiv
  \int_{0}^{X} \rd k_L
  \int_{0}^{Y} \rd k_R \
  s(k_L, k_R,\mu) .
\end{align}
%----------------------------------------------------------------------------------------------------
This function contains the entire soft contribution to the integrated doubly differential hemisphere mass distribution.
Since it is a function, rather than a distribution, we can use this integrated form to check the $\mu$-dependence
and compare to previous predictions.

We can calculate $\cR(X,Y,\mu)$ using Eq.~\eqref{result:summ} and the expansion in Eq.~\eqref{stardist}. 
For the same direction contribution (the real emission graphs, real/virtual interference graphs, and charge renormalization),
 the soft function is trivial to integrate in $d$ dimensions. 
For the opposite direction contribution, the integral
of the distributions is complicated by the overlapping singularities.
 It is a straightforward exercise in sector decomposition~\cite{Heinrich:2008si}
to isolate the singularities and perform the integrations. 
The result can then be renormalized in $\overline{\text{MS}}$. We find, for the opposite direction contribution,
%----------------------------------------------------------------------------------------------------
\begin{multline}
\cR(X,Y,\mu) = \left(\frac{\alpha_s}{4\pi}\right)^2 \Bigg\{
   \frac{1}{4} f^{(2)}(0)
  -\frac{1}{2} f^{(1)}(0) \ln   \frac{XY}{\mu^2} 
  +\frac{1}{2} f^{(0)}(0) \ln^2 \frac{XY}{\mu^2} 
  \\
  -\frac{1}{2} \int_0^1 \!\! \rd z\ \left[ \frac{1}{z} \right]_{+} f^{(1)}(z) 
  + \int_0^1 \!\! \rd z\ \left[ \frac{\ln z}{z} \right]_{+}      f^{(0)}(z) 
  + \ln \frac{XY}{\mu^2} \int_0^1 \!\! \rd z\ \left[ \frac{1}{z} \right]_{+} f^{(0)}(z) 
\\
  -\frac{1}{2} 
    \int_{1}^{Y/X} \!\! \rd y
    \int_{1}^{Y/X} \!\! \rd x \
    \frac{f^{(0)}(x/y) - f^{(0)}(0) }{xy}
\Bigg\},
\end{multline}
%----------------------------------------------------------------------------------------------------
where $f^{(n)}(r)$ refer to the coefficients in the expansion in Eq.~\eqref{fexp}.
The final compiled results for $\cR(X,Y,\mu)$ for the different color structures, including the same-direction
and opposite direction contributions, are given in Sec.~\ref{sec:hemi}.

The integrated soft function directly gives us the $\alpha_s^2$ soft function contribution
to the integrated order $\alpha_s^2$ heavy jet mass distribution,
%----------------------------------------------------------------------------------------------------
\begin{equation}
  R_\rho(\rho,\mu) = \frac{1}{\sigma_0}\int_0^\rho \frac{\rd \sigma}{\rd \rho'} \rd\rho' = \cR(\rho Q, \rho Q,\mu).
\end{equation}
%----------------------------------------------------------------------------------------------------
For thrust, the integrated distribution is not given in trivial way from the integrated soft function.
However, it differs from the heavy-jet mass distribution only by a single finite integral
%----------------------------------------------------------------------------------------------------
\begin{equation}
  R_\tau(\tau,\mu) =  \frac{1}{\sigma_0}\int_0^\tau \frac{\rd \sigma}{\rd \tau'} \rd\tau' 
= R_\rho(\tau,\mu) - \left(\frac{\alpha_s}{4\pi}\right)^2 \int_0^1\rd x \int_{1-x}^1\rd y \frac{f^{(0)}(x/y)}{xy}
\end{equation}
which we can now compute for the $C_F C_A$ and $C_F n_f T_F$ color structures. Adding also the $C_F^2$ terms,
which were already konwn, the result is
\begin{align}
 R_\tau(\tau,\mu)&= R_\rho(\tau,\mu) + \left(\frac{\alpha_s}{4\pi}\right)^2 \left[
 -\frac{8\pi^4}{45}C_F^2
+\left(\frac{8}{3} - 8\zeta_3 \right) C_F n_f T_F 
\right. \nn
&\left.
+ \left(32 \text{Li}_4\frac{1}{2}+22 \zeta_3+28 \zeta_3 \ln2-\frac{4}{3}-\frac{38 \pi^4}{45}
+\frac{4 \ln^42}{3}-\frac{4}{3} \pi ^2 \ln ^2 2\right)
 C_F C_A \right].
\end{align}
%----------------------------------------------------------------------------------------------------

%%%%%  x-x-x-x-x-x-x-x-x-x-x-x-x-x-x-x-x-x-x-x-x-x-x-x-x-x-x-x-x-x-x-x-x-x-x-x-x-x-x-x-x-x-x-x-
%%%%%  x-x-x-x-x-x-x-x-x-x-x-x-x-x-x-x-x-x-x-x-x-x-x-x-x-x-x-x-x-x-x-x-x-x-x-x-x-x-x-x-x-x-x-x-
%%%%%  x-x-x-x-x-x-x-x-x-x-x-x-x-x-x-x-x-x-x-x-x-x-x-x-x-x-x-x-x-x-x-x-x-x-x-x-x-x-x-x-x-x-x-x-

\section{Numerical check for thrust and heavy jet mass \label{sec:thrust}}
As a check on our results, we can use the soft function to calculate the soft contribution to the differential thrust and
heavy jet mass distributions. The singular parts of these distributions at $\mathcal{O}(\alpha_s^2)$ were previously determined up to four numbers: the coefficients of $\delta(\tau)$ and $\delta(\rho)$ for the $C_F C_A$ and $C_F n_f T_F$ color structures. Until now these four numbers were unknown and had to be fit numerically using the {\sc event 2} program. We can now use our results for the hemisphere soft function to replace these numerically fit numbers with analytical results. The coefficients of the $\delta$-functions are the same as the constant terms in $R_\rho(\rho)$ and $R_\tau(\tau)$, for which formulae were given in the previous section.

The unknown soft contributions to the coefficients of $\delta(\tau)$ and $\delta(\rho)$ were denoted $c_2^S$ and $c_{2\rho}^S$
in~\cite{Chien:2010kc}. We find
%-----------------------------------------------------------------------------
\begin{align}
\ctwo \label{ctwoeq}
&= 
  \frac{\pi^4}{2}
  C_F^2
  + 
  \left(
    -\frac{2140}{81}
    -\frac{871 \pi ^2}{54}
    +\frac{14 \pi^4}{15}
    +\frac{286 \zeta_3}{9}
    \right)
  C_F C_A
\nn & \qquad
  +
  \left(
     \frac{80}{81}
    +\frac{154 \pi ^2}{27}
    -\frac{104 \zeta_3}{9}
  \right)
  C_F n_f T_F,
\\
\ctwor  \label{ctworeq}
&= 
  \frac{\pi^4}{2}
  C_F^2
  + 
  \left(
    -\frac{2032}{81}
    -\frac{871 \pi^2}{54}
    +\frac{16 \pi^4}{9}
    -\frac{4 \ln^42}{3}
    +\frac{4}{3} \pi^2 \ln^2 2
    -28 \zeta_3 \ln 2
\right.
\nn 
& \qquad
\left.
    +\frac{88 \zeta_3}{9}
    -32 \text{Li}_4\left(\frac{1}{2}\right)
    \right)
  C_F C_A
  +
  \left(
    -\frac{136}{81}
    +\frac{154 \pi ^2}{27}
    -\frac{32 \zeta_3}{9}
  \right)
  C_F n_f T_F.
\end{align}
%-----------------------------------------------------------------------------

These numbers were fit numerically in~\cite{Becher:2008cf,Chien:2010kc,Hoang:2008fs} based on 
a method introduced in~\cite{Becher:2008cf}. The procedure involves subtracting the singular parts
of the thrust and heavy jet mass distributions, which are known analytically from SCET, up to delta-function terms, from the
full QCD distributions for thrust and heavy jet mass calculated numerically with the program {\sc event 2}. The difference
is then integrated over and compared to the total cross section, which is known analytically, minus the analytic integral
over the singular terms. The highest precision fits were done in~\cite{Chien:2010kc} so we compare only to those. The
result is
%-----------------------------------------------------------------------------
\begin{align}
\ctwo
&= ( 48.7045 ) C_F^2
  + 
  ( -56.4990 ) C_F C_A
  +
  ( 43.3905 ) C_F n_f T_F  && (\text{analytic result} )
\nn
&= ( 49.1 ) C_F^2
  + 
  ( -57.8 ) C_F C_A
  +
  ( 43.4 ) C_F n_f T_F    && (\text{fit result}~\cite{Chien:2010kc} )
\end{align}
%-----------------------------------------------------------------------------
and
$c_{2\rho}^{S}$
%-----------------------------------------------------------------------------
\begin{align}
\ctwor 
&= ( 48.7045 ) C_F^2
  + 
  ( -33.2286 ) C_F C_A
  +
  ( 50.3403 ) C_F n_f T_F  && (\text{analytic result} )
\nn
&= ( 49.1 ) C_F^2
  + 
  ( -33.2) C_F C_A
  +
  ( 50.2 ) C_F n_f T_F      && (\text{fit result}~\cite{Chien:2010kc} ).
\end{align}
%-----------------------------------------------------------------------------
The percent errors for these numbers are 0.8\%, 2\%, 0.02\% for $\ctwo$ and 0.8\%, 0.08\% and 0.2\% for $\ctwor$ 
respectively, with an average error of around 0.5\%. This is excellent agreement. Note that the $C_F^2$ terms
were already known when the fits were done, so small errors were expected.

For completeness, the complete contributions of the soft function to
$\delta(\rho)$ and $\delta(\tau)$, denoted by $D^{(\rho)}_{\delta}$ and $D^{(\tau)}_{\delta}$ at order $\alpha_s^2$ are
%%-----------------------------------------------------------------------------
\begin{align}
  D^{(\tau)}_\delta &=
\left(\frac{\alpha_s}{4\pi}\right)^2 \Bigg\{
c^S_2 - \frac{4}{5}\pi^4 C^2_F + C_F C_A \left( \frac{352\zeta_3}{9}
+ \frac{268\pi^2}{27} - \frac{4\pi^4}{9}\right)
%\nn
%&&
+ C_F T_F n_f \left(
-\frac{128\zeta_3}{9} - \frac{80\pi^2}{27} \right)\Bigg\} \non,
\\
D^{(\rho)}_\delta &=
\left(\frac{\alpha_s}{4\pi}\right)^2 \Bigg\{
c^S_{2\rho} - \frac{28}{45}\pi^4 C^2_F + C_F C_A \left( \frac{352\zeta_3}{9}
+ \frac{268\pi^2}{27} - \frac{4\pi^4}{9}\right)
%\nn
%&&
+ C_F T_F n_f \left(
-\frac{128\zeta_3}{9} - \frac{80\pi^2}{27} \right)\Bigg\} .
\end{align}
%%-----------------------------------------------------------------------------
One can also use $\ctwo$ and $\ctwor$ to get the complete coefficient of $\delta(\tau)$ and $\delta(\rho)$ including
jet and hard function contributions, using Appendices C of Refs.~\cite{Becher:2008cf} and~\cite{Chien:2010kc}. 

%For completeness, the complete contributions of the soft function to
%$\delta(\rho)$ and $\delta(\tau)$, denoted by $D^{(\rho)}_{\delta}$ and $D^{(\tau)}_{\delta}$, are
%%-----------------------------------------------------------------------------
%\begin{align}
%&D^{(\rho)}_{\delta} = 
%    \frac{\pi^4}{2}
%  C_F^2
%  +
%  C_F C_A
%  \left(
%    -\frac{2032}{81}
%    -\frac{871 \pi^2}{54}
%    +\frac{16 \pi^4}{9}
%    -\frac{4\ln^4(2)}{3}
%    +\frac{4\pi^2}{3}  \ln^22
%\right.
%\nn
%& \qquad
%\left.
%    +\frac{88 \zeta_3}{9}
%    -28 \zeta_3\ln 2
%    -32 \text{Li}_4\left(\frac{1}{2}\right)
%  \right)
%  +
%  C_F n_F T_F
%  \left(
%    -\frac{136}{81}
%    +\frac{74 \pi ^2}{27}
%    -\frac{160 \zeta_3}{9}
%  \right) \non
%\end{align}
%%-----------------------------------------------------------------------------
%and
%%-----------------------------------------------------------------------------
%\begin{align}
%&D^{(\tau)}_{\delta} =
%    \frac{\pi^4}{2}
%  C_F^2
%  + 
%  C_F C_A
%  \left(
%    -\frac{2140}{81}
%    -\frac{871 \pi ^2}{54}
%    +\frac{14 \pi^4}{15}
%    +\frac{286 \zeta_3}{9}
%    \right)
%% & \qquad
%  +
%  C_F n_f T_F
%  \left(
%     \frac{80}{81}
%    +\frac{74 \pi ^2}{27}
%    -\frac{232 \zeta_3}{9}
%  \right)\non
%\end{align}
%-----------------------------------------------------------------------------

%%%%%  x-x-x-x-x-x-x-x-x-x-x-x-x-x-x-x-x-x-x-x-x-x-x-x-x-x-x-x-x-x-x-x-x-x-x-x-x-x-x-x-x-x-x-x-
%%%%%  x-x-x-x-x-x-x-x-x-x-x-x-x-x-x-x-x-x-x-x-x-x-x-x-x-x-x-x-x-x-x-x-x-x-x-x-x-x-x-x-x-x-x-x-
%%%%%  x-x-x-x-x-x-x-x-x-x-x-x-x-x-x-x-x-x-x-x-x-x-x-x-x-x-x-x-x-x-x-x-x-x-x-x-x-x-x-x-x-x-x-x-
\section{Hemisphere mass distribution \label{sec:hemi}}
The numerical check performed in the previous section provides strong evidence that our analytical results are correct.
With these results in hand, we can now compare to other features of the hemisphere mass distribution and the integrated
hemisphere soft function, $\cR(X,Y,\mu)$.

The entire $\mu$-dependence of $\cR(X,Y,\mu)$ is predicted by SCET. Indeed, renormalization group invariance predicts 
that the differential soft function must factorize in Laplace space, as in Eq.~\eqref{softfact}. The Laplace transform is defined 
by
%-----------------------------------------------------------------------------
\begin{equation}
  \tilde{s}(x_L, x_R, \mu ) = \int_0^\infty \rd k_L \int_0^\infty \rd k_R S(k_L,k_R,\mu)e^{- x_L k_L e^{-\gamma_E}} e^{- x_R k_R e^{-\gamma_E}} ,
\end{equation}
%-----------------------------------------------------------------------------
where the $\gamma_E$ factors are added in the definition to avoid their appearance elsewhere.
The factorization theorem then implies
%-----------------------------------------------------------------------------
\begin{equation} \label{sfactb}
\tilde{s}(x_L, x_R, \mu ) = \smu(\ln x_L \mu) \smu(\ln x_R \mu) \sf(x_L,x_R). 
\end{equation}
%-----------------------------------------------------------------------------
The RG-kernel $\smu(L)$ is determined by the renormalization group invariance of the
factorization formula, and is expressible in terms of the anomalous dimensions of the hard and
jet functions, which are known up to $\alpha_s^3$. The finite part $\sf(L)$, until now, has been known only to $\alpha_s$.
This Laplace form leads to a simple expression for the integrated soft function in SCET~\cite{Chien:2010kc}
%-----------------------------------------------------------------------------
\begin{equation} 
\cR(X,Y,\mu) = \tilde{s}(\partial_{\eta_1}, \partial_{\eta_2},\mu)
 \left(\frac{X}{\mu}\right)^{\eta_1}
\frac{e^{-\gamma_E\eta_1}}{\Gamma(\eta_1 +1)}
 \left(\frac{Y}{\mu}\right)^{\eta_2}
\frac{e^{-\gamma_E\eta_2}}{\Gamma(\eta_2 +1)}
\Big|_{\eta_1 =\eta_2=0}.
\end{equation}
%-----------------------------------------------------------------------------
The $\mu$-dependent terms in the order $\alpha_s^2$ integrated soft function calculated in this way agree exactly with
the $\mu$-dependent terms in $\cR(X,Y,\mu)$. In fact, it is helpful to separate out those terms. To that end, we write the
$\alpha_s^2$ terms as
%-----------------------------------------------------------------------------
\begin{equation}
  \cR(X,Y,\mu) =\left(\frac{\alpha_s}{4\pi}\right)^2\left[\cR_\mu\left(\frac{X}{\mu},\frac{Y}{\mu}\right)+ \cR_f\left(\frac{X}{Y}\right)\right],
\end{equation}
%-----------------------------------------------------------------------------
where $\cR_\mu(X/\mu, Y/\mu)$ is the part coming directly from the $\smu(L)$ terms and $\cR_f(X/Y)$ is the remainder,
which comes from $\sf(x_L,x_R)$. The result for $\cR_\mu(X/\mu, Y/\mu)$ is
%-----------------------------------------------------------------------------
\begin{align}
&\cR_\mu\left(\frac{X}{\mu},\frac{Y}{\mu}\right)
=
\Bigg[
8\ln^{4}\frac{X}{\mu}-\frac{20}{3}\pi^{2}\ln^{2}\frac{X}{\mu}+16\ln^{2}\frac{X}{\mu}\ln^{2}\frac{Y}{\mu}\nn &\quad+64\zeta_3\ln\frac{XY}{\mu^{2}}+8\ln^{4}\frac{Y}{\mu}-\frac{20}{3}\pi^{2}\ln^{2}\frac{Y}{\mu}-\frac{28\pi^{4}}{45}
\Bigg]C^2_F
\nn
&\quad
+\Bigg[
\frac{88}{9}\ln^{3}\frac{X}{\mu}+\frac{4}{3}\pi^{2}\ln^{2}\frac{X}{\mu}-\frac{268}{9}\ln^{2}\frac{X}{\mu}-\frac{22}{9}\pi^{2}\ln\frac{XY}{\mu^{2}}+\frac{808}{27}\ln\frac{XY}{\mu^{2}}\nn &\quad-28\zeta_3\ln\frac{XY}{\mu^{2}}+\frac{88}{9}\ln^{3}\frac{Y}{\mu}+\frac{4}{3}\pi^{2}\ln^{2}\frac{Y}{\mu}-\frac{268}{9}\ln^{2}\frac{Y}{\mu}+\frac{352\zeta_3}{9}-\frac{4\pi^{4}}{9}+\frac{268\pi^{2}}{27}
\Bigg]C_F C_A
\nn
&\quad
+\Bigg[
-\frac{32}{9} \ln ^3\frac{X}{\mu }+\frac{80}{9} \ln ^2\frac{X}{\mu
   }+\frac{8}{9} \pi ^2 \ln \frac{X Y}{\mu ^2}-\frac{224}{27} \ln
   \frac{X Y}{\mu ^2}
\nn
&~~~~~~~~~~~~~~~~~~~~~~~~~~\quad
-\frac{32}{9} \ln ^3\frac{Y}{\mu
   }+\frac{80}{9} \ln ^2\frac{Y}{\mu }-\frac{128 \zeta_3}{9}-\frac{80 \pi ^2}{27}
\Bigg]C_F T_F n_f.
\end{align}
%-----------------------------------------------------------------------------

The part of the soft function not determined by RG-invariance is represented entirely by $\sf(x_L,x_R)$. This function
is $\mu$-independent and can only depend on the ratio $x_L/x_R$ by dimensional analysis. Moreover, it is
symmetric in $x_L \leftrightarrow x_R$, since the hemisphere soft function is symmetric in $k_L \leftrightarrow k_R$. 
Hoang and Kluth claimed~\cite{Hoang:2008fs} that it should only have logarithms,
and up to order $\alpha_s^2$,  only have $\ln^0$ and $\ln^2$ terms. Their ansatz was that
%-----------------------------------------------------------------------------
\begin{equation}
\sf(x_L,x_R)^{\text{Hoang-Kluth}} =
1 
+ \left( \frac{\alpha_s}{4\pi} \right) c_1^{S} 
+ \left( \frac{\alpha_s}{4\pi} \right)^2[\ctwo + \ctwoL \ln^2\frac{x_L}{x_R} ],
\end{equation}
%-----------------------------------------------------------------------------
with $\cone=-C_F\pi^2$ already known.

To check the Hoang-Kluth ansatz, the easiest approach is to look at the contribution of $\sf(x_L,x_R)$ to $\cR(X,Y,\mu)$,
which we called $\cR_f(X/Y)$.  For the Hoang-Kluth ansatz, the result is
%-----------------------------------------------------------------------------
\begin{equation}
  \cR_f(z)^{\text{Hoang-Kluth}} = \ctwo+ \ctwoL( \ln^2 z - \frac{\pi^2}{3}) .
\end{equation}
%-----------------------------------------------------------------------------
The values of $\ctwo$ and $\ctwoL$ which get right the singular parts of the thrust and heavy jet mass distributions
are given in Eqs.~\eqref{ctwoeq} and \eqref{ctworeq} with $\ctwoL = \frac{3}{\pi^2}(\ctwo - \ctwor)$.

The exact answer, at order $\alpha_s^2$ is
%-----------------------------------------------------------------------------
\begin{align}
\label{sf:exact}
&\cR_f(z) =  
%\left(\frac{\alpha_s}{4\pi}\right)^2 \Bigg\{
  \frac{\pi^4}{2}
  C_F^2
  +
\left[
   -88 \text{Li}_3(-z)-16 \text{Li}_4\left(\frac{1}{z+1}\right)-16
   \text{Li}_4\left(\frac{z}{z+1}\right)+16 \text{Li}_3(-z) \ln
   (z+1)
\right.
\nn
&\qquad 
\left.
   +\frac{88 \text{Li}_2(-z) \ln (z)}{3}-8 \text{Li}_3(-z) \ln
   (z)-16 \zeta (3) \ln (z+1)+8 \zeta (3) \ln (z)-\frac{4}{3} \ln^4(z+1)
\right.
\nn
&\qquad 
\left.
   +\frac{8}{3} \ln (z) \ln^3(z+1)+\frac{4}{3} \pi^2 \ln^2(z+1)-\frac{4}{3} \pi^2 \ln^2(z)-\frac{4 \left(3 (z-1)+11 \pi^2 (z+1)\right) \ln (z)}{9 (z+1)}
\right.
\nn
&\qquad 
\left.
    -\frac{506 \zeta (3)}{9}+\frac{16\pi^4}{9}-\frac{871 \pi^2}{54}-\frac{2032}{81}\right]C_F C_A
+
  \left[ 
  32 \text{Li}_3(-z)-\frac{32}{3} \text{Li}_2(-z) \ln (z)
\right.
\nn
&\qquad 
\left.
    +\frac{8 (z-1) \ln (z)}{3 (z+1)}+\frac{16}{9} \pi^2 \ln (z)+\frac{184 \zeta (3)}{9}+\frac{154 \pi^2}{27}-\frac{136}{81}
  \right]
  C_F n_f T_F
%  \Bigg\}.
\end{align}
%-----------------------------------------------------------------------------
This is clearly very different from the Hoang-Kluth form.

%%%%%  x-x-x-x-x-x-x-x-x-x-x-x-x-x-x-x-x-x-x-x-x-x-x-x-x-x-x-x-x-x-x-x-x-x-x-x-x-x-x-x-x-x-x-x-
%%%%%  x-x-x-x-x-x-x-x-x-x-x-x-x-x-x-x-x-x-x-x-x-x-x-x-x-x-x-x-x-x-x-x-x-x-x-x-x-x-x-x-x-x-x-x-
%%%%%  x-x-x-x-x-x-x-x-x-x-x-x-x-x-x-x-x-x-x-x-x-x-x-x-x-x-x-x-x-x-x-x-x-x-x-x-x-x-x-x-x-x-x-x-

\section{Asymptotic behavior and non-global logs \label{sec:asym}}
The factorization theorem is valid in the dijet limit when the hemisphere masses are small compared
to $Q$; however, there is no restriction on the relative size of the two masses.  
In addition to logarithms $\ln \frac{M_{L,R} }{\mu}$ required by RG invariance, there may be
logarithms of the form $\ln \frac{M_{L}}{M_{R}}$ that enter at order $\alpha_s^2$.
 These logarithms cannot be predicted by RG invariance and are known as
non-global logarithms.
Salam and Dasgupta have shown that non-global logs appear in distributions such as the light jet mass.
They argued that in the strongly-ordered soft limit, when $M_L \ll M_R \ll Q$,  
the leading non-global log should be $-(\frac{\alpha_s}{4\pi})^2\frac{4\pi^2}{3} 
C_F C_A \ln^2\frac{M_L^2}{M_R^2}$ in full QCD. This double log was reproduced
%, also in the strongly-ordered soft limit,  but 
%starting from Eikonal Feynman rules directly, 
in~\cite{scetCL}.

Non-global logs must be present in SCET, since for small $M_L$ and $M_R$, the entire distribution
is determined by soft and collinear degrees of freedom.
% However, SCET may not be able to resum these logs. 
The non-global logs cannot come from the hard function, which has no knowledge of either mass, or the jet
function, since each jet function knows about only one mass. Thus, they must come from the soft function.
Moreover since, by definition, they are not determined by RG invariance, they must be present in the $\mu$-independent
part, $\cR_f(X/Y)$ of the integrated hemisphere soft function, $\cR(X,Y,\mu)$. This function was given explicitly in Eq.~\eqref{sf:exact}.

To see the non-global logs in $\cR_f(z)$ we can simply take the limit $z \to \infty$. Note that $\cR_f(z)=\cR_f(\frac{1}{z})$ so this
is also the limit $z\to 0$. The asymptotic limit of $\cR_f(z)$ for large or small $z$ is
%-----------------------------------------------------------------------------
\begin{align}
\label{sf:large}
&\cR_f^{z\gg 1}(z) =
  \frac{\pi^4}{2}
  C_F^2
+
  \left[ 
  \left(
    \frac{8}{3} - \frac{16\pi^2}{9} 
  \right) |\ln z|
  +
  -\frac{136}{81}
  +\frac{154 \pi ^2}{27}  
  +
  \frac{184 \zeta_3}{9}
  \right]
  C_F n_f T_F \\
&+  \left[ 
    -\frac{4}{3} \pi ^2 \ln^2 z
    +
      \left(
        -8 \zeta_3
        -\frac{4}{3}
        +\frac{44 \pi ^2}{9}
      \right)
      |\ln z|
   -\frac{506 \zeta_3}{9}
%\right.
%\nn
%&\qquad 
%\left.
   +\frac{8 \pi^4}{5}
   -\frac{871 \pi ^2}{54}
   -\frac{2032}{81}
  \right]
  C_F C_A  \non.
\end{align}
%-----------------------------------------------------------------------------
There are two important features to note in this expansion.
 First of all, in the $C_F C_A$ color structure there is a term $-\frac{4\pi^2}{3} \ln^2 z$,
which is the leading non-global log found by Dasgupta and Salam and in~\cite{scetCL}.
But we also see that there are sub-leading non-global logs, of
the form $|\ln z|$. The absolute value is necessary to keep the expression 
symmetric in $z\to \frac{1}{z}$. It is interesting to see how this sign flip comes out of the full analytic expression. 

%----------FIGURE-------------------------------------------------------------
\begin{figure}[t]
\begin{center}
\includegraphics[width=0.45\textwidth]{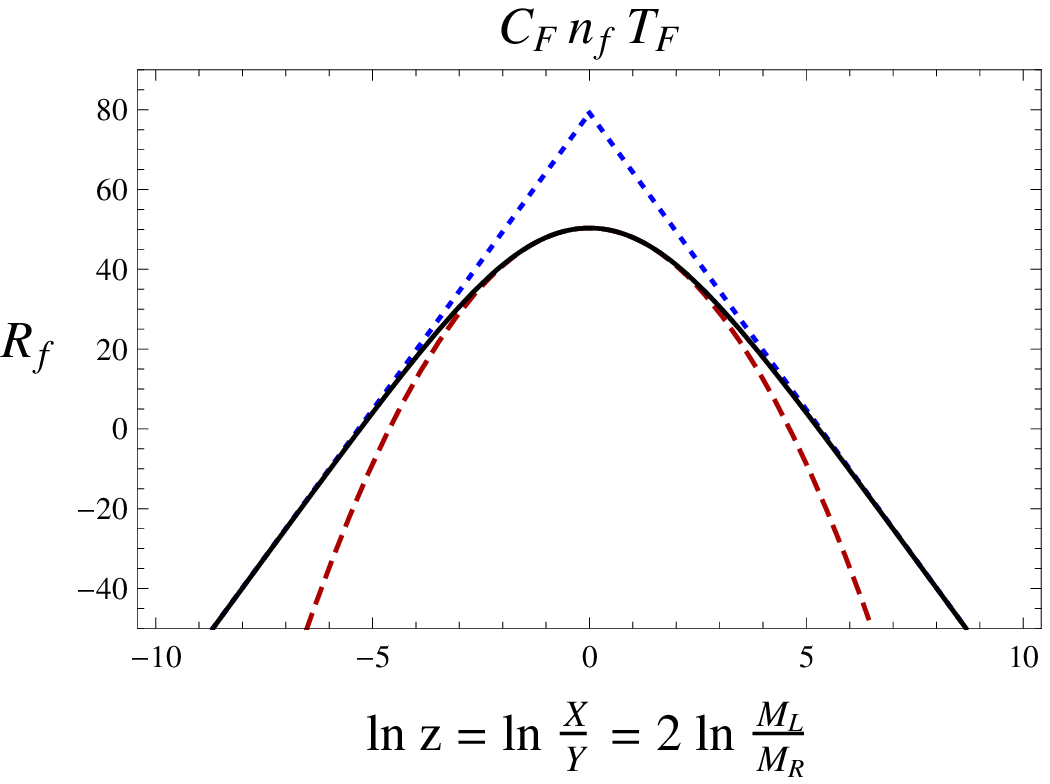}
\hspace{0.5cm}
\includegraphics[width=0.45\textwidth]{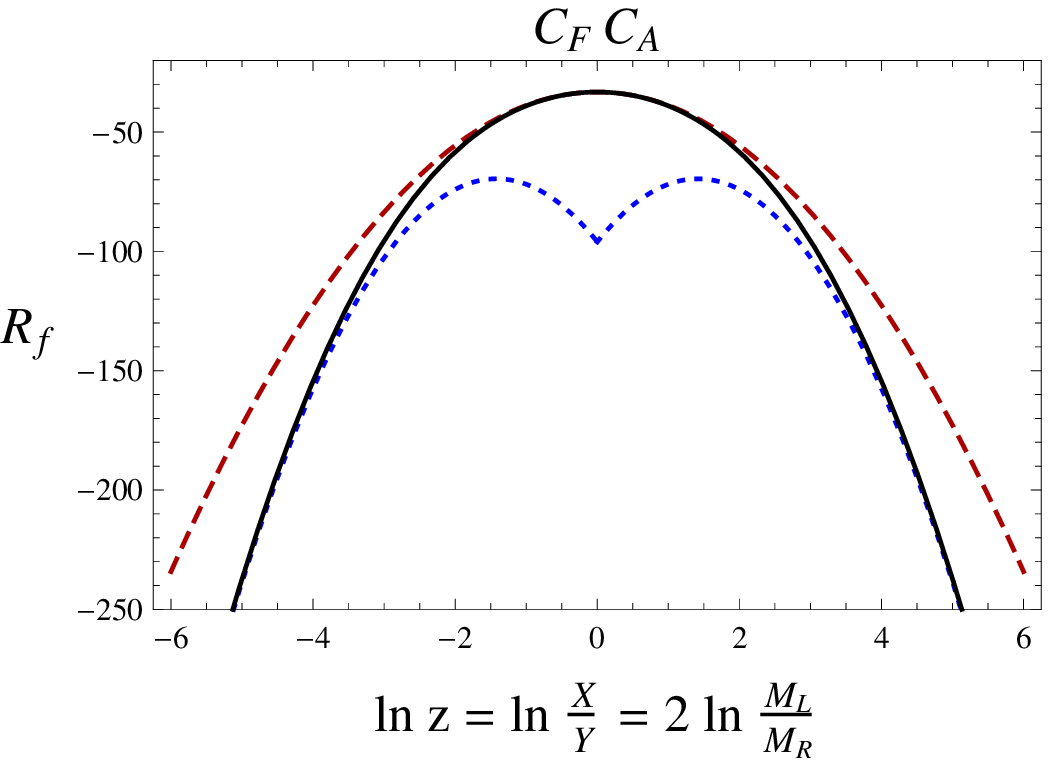}
\end{center}
\caption{The contribution of the part of the soft function not fixed by renormalization group invariance
to the hemisphere mass distribution, $\cR_f(z)$ is shown. On the left is the  $C_F n_f T_F$ 
color factor and on the right is $C_F C_A$, both as a function of
$\ln z \equiv \ln \frac{X}{Y}$.  The solid black curve
is the exact result of Eq.~(\ref{sf:exact}).  The dashed red curve is the plot of the small $\ln z$
expression of Eq.~(\ref{sf:small}) and the dotted blue curve give the large $\ln z$ behavior
of Eq.~(\ref{sf:large}). The kink is due to a sign flip, since the linear term appears as $|\ln z|$.}
\label{fig:sf}
\end{figure}
%-----------------------------------------------------------------------------

Next, let us look at $z\sim 1$.
Here we find
%-----------------------------------------------------------------------------
\begin{align}
\label{sf:small}
&\cR_f^{z\sim1}(z) = 
  \frac{\pi^4}{2}
  C_F^2
  +
  \left[ 
    \left(
      -\frac{2}{3}
      -\frac{4 \pi ^2}{3}
      -4 \ln ^2 2
      +\frac{44 \ln  2}{3}
    \right)
\ln^2z
    -32 \text{Li}_4\left(\frac{1}{2}\right)
    +\frac{88 \zeta_3}{9}
\right.
\nn
&\qquad 
\left.
    -28 \zeta_3 \ln (2)
    -\frac{2032}{81}-\frac{871 \pi^2}{54}
    +\frac{16 \pi ^4}{9}
    -\frac{4 \ln ^4 2}{3}
    +\frac{4}{3} \pi ^2 \ln ^2 2
  \right]
  C_F C_A  
\nn
&\qquad 
+
  \left[ 
    \left(
      \frac{4}{3}
      -\frac{16 \ln 2}{3}
    \right)
\ln^2z
    +\frac{154 \pi ^2}{27}
    -\frac{136}{81}
    -\frac{32 \zeta_3}{9}
  \right]
  C_F n_f T_F  + \mathcal{O}(\ln^3 z).
\end{align}
%-----------------------------------------------------------------------------
We see there is a double logarithmic term for both the $C_F C_A$ and $C_F n_f T_F$ color structures.
 This is consistent with an analysis performed in~\cite{Hoang:2008fs} of
an observable $\rho^\alpha = \max(\alpha M_L^2,M_R^2)/Q^2$. They found that the integrated $\rho^\alpha$ distribution
looked like $\ln^2\alpha$ for $\alpha\sim 1$. This quadratic behavior in $\ln \alpha$ corresponds exactly to the quadratic behavior
in the  $z\sim 1$ limit in~\eqref{sf:small}.

We show in Figure~\ref{fig:sf}
the exact finite function $\cR_f(z)$ and a comparison to its asymptotic behavior at small $\ln z$ and large $\ln z$
for the $C_F C_A$ and $C_F n_f T_F$ color structures.
For both color structures, the exact curve is well approximated by a parabola for small $\ln z$.
At large $\ln z$, for the $C_F n_f T_F$ color factor, the exact result approaches a linear function whereas  
the $C_F C_A$ color structure has $\ln^2 z$ dependence with a different coefficient
than for the small $\ln z$ limit.  The $C_F C_A$ term has a linear term as well. 

As we have discussed, the integrated hemisphere soft function contributes directly to the doubly
differential hemisphere mass distribution. In the limit where both hemisphere masses are small,
and well separated, the soft function gives the dominant contribution. In this regime, 
we can read off that the leading non-global logarithms are given by $\cR_f^{z\gg 1}(M_L^2/M_R^2)$ in
Eq.\eqref{sf:large}. The $\ln^2$ term has an identical coefficient to that
found in~\cite{Dasgupta:2001sh}. The subleading non-global logarithm is a new result.
%%-----------------------------------------------------------------------------
%\begin{align}
%\label{ngl}
%R(M_L,M_R) &= \left(\frac{\alpha_s}{4\pi}\right)^2\left\{  -\frac{4\pi^2}{3} 
%\ln^2 \frac{M_L^2}{M_R^2}  C_F C_A    +
%      \left(
%        -8 \zeta_3
%        -\frac{4}{3}
%        +\frac{44 \pi ^2}{9}
%      \right)\left|\ln \frac{M_L^2}{M_R^2}\right|
%C_F C_A
%\right. \nn
%&
%\left.
% + \left|\ln \frac{M_L^2}{M_R^2}\right|
%  \left(
%    \frac{8}{3} - \frac{16\pi^2}{9} 
%  \right) C_F n_f T_F
%\right\},
%\end{align}
%-----------------------------------------------------------------------------

\section{Exponentiation \label{sec:exp}}
These non-global logarithms become important when one scale becomes parametrically larger than the other.
The separation of scales suggest that at the higher of the two scales, one may be able to match onto 
a new effective theory and then run the matching coefficient between the two scales. 
In fact, the $-\frac{4 \pi^2}{3} \ln^2 z$ term in this calculation has its origin in 
$f(0,0)$, where $f(r,\e)$ is the opposite-direction contribution to the hemisphere soft function,
as in the Appendix. Indeed, we found that 

%Then it may
%be possible to resum at least the leading logarithmic dependence.  Note that the only place where the two 
%scales appear is when the two partons propagate into separate hemisphere which resulted in a dependence
%on the function $f(r)$.  When matching, only the leading terms are kept
%in the power counting, in this case $r = k_L/k_R$, and so the leading matching coefficient would
%be proportional to $f(0)$.  Indeed, for our calculation
%-----------------------------------------------------------------------------
\begin{equation}
f_{C_A}(0,\e)  = \frac{8\pi^2}{3} + \mathcal{O}(\e), \qquad
 f_{nf}(0,\e) = \mathcal{O}(\e^2),
\end{equation}
%-----------------------------------------------------------------------------
which is consistent with the leading non-global logarithm only having the $C_F C_A$ color structure.
Since it is the $\e^0$ part of this expression which contributes, and there are double soft poles,
the full expansion also has terms like $f(0,0)\ln^2\mu$. Thus $f(0,0)$ can be thought of as an anomalous dimension,
providing hope that these non-global logs might be resummed in an effective theory. A consistent framework
may require some kind of refactorization, like the one found for a related event shape, $\tau_\omega$, in~\cite{Kelley:2011tj}.
Ideas along these lines were suggested in talks by Chris Lee~\cite{scetCL,boostCL}. Lee and
collaborators proposed that the leading non-global logs might be resummed with effective field theory,
although no details were given.

There is actually good reason to believe the resummation of non-global logs is more challenging than the types of
resummation done in SCET. To see this, we first consider the predictions from non-Abelian exponentiation. Non-Abelian
exponentiation applies only to the case of pure QCD, without fermion loops. In this case, it says that the full
soft function, in Laplace space, can be written as an exponential of 2-particle irreducible diagrams. At order
$\alpha_s^n$, new contributions can appear only to the maximally non-Abelian color structure, $C_F C_A^{n-1}$. For example,
at two-loops, this tells us that the $C_F^2$ color structure is given entirely by the exponential of the one-loop
$C_F$ color structure. At 3-loops it predicts the entire $C_F^3$ and $C_F^2 C_A$ color structures.

 To be more specific, the soft function in Laplace space  factorizes as in Eq.~\eqref{sfactb}, with the $\smu$ terms
and the $\sf$ terms separately exponentiating, as explained in~\cite{Hoang:2008fs}. So we can write
\begin{equation}
\sf(x_L,x_R) = \exp \left[\frac{\alpha_s}{4\pi}( -\pi^2)C_F 
+ \left(\frac{\alpha_s}{4\pi}\right)^2 \left( C_F n_f T_F\sf^{(2,n_f)}(x_L,x_R)+  C_FC_A\sf^{(2,C_A)}(x_L,x_R)  \right) + \cdots \right]    
\label{naeform}
\end{equation}
where $\sf^{(2,n_f)}$ and $\sf^{(2,C_A)}$ are the Laplace transforms of the $C_F n_f T_F$ and $C_F C_A$ color structures
in the 2-loop soft function.\footnote{Although we have not computed the Laplace-space soft function
directly, it was calculated by another group after the first version
of this paper appeared~\cite{Hornig:2011iu}. It has a qualitatively similar form to $\cR_f(\frac{x_L}{x_R})$.}
Such a rewriting has no content unless there is some restriction on the terms appearing in the exponent. Non-Abelian
exponentiation tells us that the higher-order terms with $C_F$ and $C_A$'s only must be maximally non-Abelian, $C_F C_A^{n-1}$.

This implies, for example, that at 3-loops we know 2 color structures. Explicitly,
\begin{equation}
  \sf^{~\text{3-loop}}(x_L,x_R) = \left(\frac{\alpha_s}{4\pi}\right)^3\left[ C_F^3 \frac{(-\pi^2)^3}{6} +  C_F^2 C_A (-\pi^2)
\sf^{(2,C_A)}(x_L,x_R) + \cdots  \right]
\end{equation}
There are 4 remaining color structures, $C_F C_A^2$, $C_F n_f^2 T_F^2$, $C_F C_A  n_f T_F$ and $C_F^2 n_f T_F$ which are still unknown.
Actually, the $C_F n_f^2 T_F^2$ color structure at 3-loops should not be hard to compute,
 but there is no known general formula for how the $n_f$ color structures exponentiate 
(see~\cite{Berger:2002sv} for some discussion).

 From the exponentiation formula,  one can read off the missing
parts of the soft contribution to the 3-loop thrust and heavy-jet mass distributions. Indeed, for $n\ge 2$, we have
\begin{equation}
  c^S_n =C_F^n  \frac{(-\pi^2)^n}{n!}  +C_F^{n-1} C_A \frac{(-\pi^2)^{n-2}}{(n-2)!} \left[ \ctwo \Big| _{C_F C_A} \right] +\cdots \,,
\end{equation}
and similarly for $c^2_{n\rho}$, 
with $\ctwo$ and $\ctwor$ given in Eqs.~\eqref{ctwoeq} and \eqref{ctworeq}. 
%\begin{equation}
%  c^S_{n\rho} =C_F^n  \frac{(-\pi^2)^n}{n!}  +C_F^{n-1} C_A \frac{(-\pi^2)^{n-2}}{(n-2)!} \left[ \ctwor \Big| _{C_F C_A} \right] \,,
%\end{equation}
%with $\ctwo$ and $\ctwor$ given in Eqs.~\eqref{ctwoeq} and \eqref{ctworeq}. 
These constants can be included in future $\alpha_s$ fits or, once the finite part of the 3-loop jet function is computed,
compared to extractions from the full thrust distrubtion at NNLO~\cite{Monni:2011gb}.

Returning to the exponentiation of non-global logs, recall that the leading non-global log comes from
$\sf^{(2,C_A)}(x_L,x_R) = -\frac{4\pi^2}{3} L^2$, with $L= \ln\frac{ x_L}{x_R}$. Thus non-Abelian exponentiation
predicts a series with terms $(C_F C_A \alpha_s^2 L^2)^{n-1}$, as well as cross-terms with the $C_F$ one-loop color structure
which are subleading. The question is whether this is the entire resummation of the leading non-global log. It
seems like the answer is no, since there is no apparent reason why 3-loop graphs cannot produce terms which scale
like $\alpha_s^3 C_F C_A^2 L^3$ (or even $\alpha_s^3 L^4$) for large $L$. A clue that these terms do exist comes
from the numerical resummation of the leading non-global log at large $N_c$ in~\cite{Dasgupta:2001sh}. These authors found that
the resummed distribution could be fit by an exponential, but it is numerically different from the pure $(C_F C_A \alpha_s^2 L^2)^{n-1}$
terms predicted by Eq.~\eqref{naeform}. Since $C_F$ and $C_A$ both scale as $N_c$ at large $N_c$, this implies that there
must be a $\alpha_s^3 C_F C_A^2 L^3$ term at 3-loops (and no $\alpha_s^3 L^4$ term). Thus the resummation of even the leading
non-global log may require a way to predict arbitrarily complicated color structures. It would be exciting to see how
this can be done in the effective field theory framework.

%%%%%  x-x-x-x-x-x-x-x-x-x-x-x-x-x-x-x-x-x-x-x-x-x-x-x-x-x-x-x-x-x-x-x-x-x-x-x-x-x-x-x-x-x-x-x-
%%%%%  x-x-x-x-x-x-x-x-x-x-x-x-x-x-x-x-x-x-x-x-x-x-x-x-x-x-x-x-x-x-x-x-x-x-x-x-x-x-x-x-x-x-x-x-
%%%%%  x-x-x-x-x-x-x-x-x-x-x-x-x-x-x-x-x-x-x-x-x-x-x-x-x-x-x-x-x-x-x-x-x-x-x-x-x-x-x-x-x-x-x-x-
\section{Conclusions \label{sec:conc}}
In this paper, we have presented the complete calculation of the hemisphere soft function to order $\alpha_s^2$. 
This is the first 2-loop calculation of a soft function which depends on two scales in addition to the
renormalization group scale $\mu$.  The hemisphere soft function, $S(k_L,k_R,\mu)$, depends on the components of the momenta going into the left and right hemispheres.
In a one-scale soft function, such as the Drell-Yan soft function, 
$S_{DY}(k,\mu)$~\cite{Korchemsky:1993uz, Belitsky:1998tc, Becher:2007ty}, the thrust soft function 
$S_T(k,\mu)$~\cite{Schwartz:2007ib,Becher:2008cf}
or the direct photon soft function $S_{\gamma}(k,\mu)$~\cite{Becher:2009th},
all of the $k$ dependence is fixed once the $\mu$-dependence is known.
Since the $\mu$-dependence is fixed by RG invariance, these functions are often completely determined. 
For multi-scale soft functions, like the hemisphere soft function,
there can be additional dependence on the ratio $r= k_L/k_R$.  We worked out this
dependence explicitly at order $\alpha_s^2$, and the result is more complicated than previously anticipated.

We performed a number of checks on our calculation. The $\mu$-dependence of the result was entirely known 
by virtue of the factorization theorem in SCET, and we have confirmed that the $\mu$-dependence of our 
hemisphere soft function matches the result obtained from factorization analysis. 
In addition, the result allows us to produce analytic expressions for all of the singular terms in
the 2-loop thrust and heavy jet mass distributions. 
The constant terms in the singular distributions were previously
unknown and had to be extracted from numerical fits~\cite{Becher:2008cf,Hoang:2008fs,Chien:2010kc}. 
We found our analytical results to be in excellent agreement with the very precise recent numerical fit of~\cite{Chien:2010kc}.

The full hemisphere soft function produces the leading and sub-leading non-global logs
in the hemisphere mass distribution. Previously,
only the leading double-log term was known, from a calculation in the soft limit of full QCD~\cite{Dasgupta:2001sh}.
In this work we reproduced that double logarithm
and, furthermore, showed the existence of a sub-leading single logarithm. This single logarithm, of $M_L/M_R$,
is interesting because $\ln M_L/M_R$ seems like it should be forbidden by the $M_L \leftrightarrow M_R$ symmetry. Curiously, we find that the complicated behavior of the hemisphere mass distribution when $M_L \sim M_R$
allows the single log to flip sign and it manifests itself as $\ln[ \max(M_L,M_R)/\min(M_L,M_R)] = |\ln M_L/M_R|$. 
Our calculation is the first to exhibit a sub-leading non-global logarithm of this type.

Besides being of formal interest, the hemisphere soft function at $\mathcal{O}(\alpha_s^2)$ is a crucial component of the resummed heavy-jet mass distribution at N$^3$LL order. Previous fits to $\alpha_s$ at this order assumed a simple form for the soft function, using the Hoang-Kluth ansatz. We have shown that this ansatz is valid only in the limit that $k_L \sim k_R$.
With the exact $\mathcal{O}(\alpha_s^2)$ soft function in hand, one source of uncertainty in the $\alpha_s$ fits to event shapes can be removed.

This work also has implications for calculations of distributions at hadron colliders. At hadron colliders, there are
necessarily many more scales in relevant observables than at $e^+e^-$ machines. For example, jet sizes and veto scales play
a critical role in many analysis~\cite{Ellis:2009wj, Ellis:2010rw,  Kelley:2011tj}. For multi-scale observables to be computed in effective field theory, we need a better
understanding of multi-scale soft functions, such as this exact result on the 2-loop hemisphere soft function provides.

\section{Acknowledgements}
The authors would like to thank Y.-T.~Chien, M.~Dasgupta, C.~Lee, K.~Melnikov, G.~Salam
 and I.~Stewart for useful discussions and F.~Petriello and I.~Scimemi
for collaboration on intermediate stages of this project.
RK and MDS were supported in part by the Department of Energy, under grant DE-SC003916.
HXZ was supported by the National Natural Science Foundation of China under grants No.~11021092 and No.~10975004
and the Graduate Student academic exchange program of Peking University.

%%%%%  x-x-x-x-x-x-x-x-x-x-x-x-x-x-x-x-x-x-x-x-x-x-x-x-x-x-x-x-x-x-x-x-x-x-x-x-x-x-x-x-x-x-x-x-
%%%%%  x-x-x-x-x-x-x-x-x-x-x-x-x-x-x-x-x-x-x-x-x-x-x-x-x-x-x-x-x-x-x-x-x-x-x-x-x-x-x-x-x-x-x-x-
%%%%%  x-x-x-x-x-x-x-x-x-x-x-x-x-x-x-x-x-x-x-x-x-x-x-x-x-x-x-x-x-x-x-x-x-x-x-x-x-x-x-x-x-x-x-x-

\appendix
\section{ Opposite direction contributions}
%Expansions of the soft function}
%\section{Appendix
 \label{app:f} 

The following are the first 3 terms in the $\e$ expansion of $f_{C_A}$ from 
Eq.~(\ref{fca:exp}).
%----------------------------------------------------------------------------------------------------
\begin{align}
f_{C_A}^{(0)}(r)
&= 
8 
\left(
  \frac{  r \left(11 r^2+21 r+12 \right) \ln (r)}{3 (r+1)^3}
    +\frac{\pi^2 (r+1)^2+2 r}{3 (r+1)^2}
    +\ln ^2(r+1)
\right.
\nn
&\qquad
\left.
    -\ln (r) \ln (r+1)
    -\frac{11}{3} \ln(r+1)
\right),
\end{align}
%----------------------------------------------------------------------------------------------------
%----------------------------------------------------------------------------------------------------
\begin{align}
f_{C_A}^{(1)}(r)
&= 
  \frac{8 \left(-11 r^3-9 r^2+9 r+11\right) \text{Li}_2(-r)}{3
   (r+1)^3}+24 \text{Li}_3(-r)-16 \text{Li}_2(-r) \log (r)
\nn
&
 +\frac{4 r
   \left(11 r^2+21 r+12\right) \log^2(r)}{3 (r+1)^3}-\frac{8 r
   \left(67 r^2+141 r+60\right) \log (r)}{9 (r+1)^3}
\nn
&
  -\frac{4 \left(r^3
   \left(11 \pi^2-36 \zeta (3)\right)+r^2 \left(-108 \zeta (3)+32+21
   \pi^2\right)+4 r \left(-27 \zeta (3)+8+3 \pi^2\right)-36 \zeta
   (3)\right)}{9 (r+1)^3}
\nn
&
  +\frac{8 \left(-11 r^3-9 r^2+9 r+11\right)
   \ln (r+1) \ln (r)}{3 (r+1)^3}-4 \ln (r+1) \ln^2(r)+\frac{4}{9}
   \left(134+3 \pi^2\right) \ln (r+1),
\end{align}
%----------------------------------------------------------------------------------------------------
%----------------------------------------------------------------------------------------------------
\begin{align}
f_{C_A}^{(2)}(r)
&=
\frac{8 \left(67 r^3+81 r^2-81 r-67\right) \text{Li}_2(-r)}{9
   (r+1)^3}-\frac{8 \left(55 r^3+117 r^2+81 r+11\right)
   \text{Li}_3(-r)}{3 (r+1)^3}
\nn
&
 -\frac{32 \left(11 r^3+9 r^2-9
   r-11\right) \text{Li}_3\left(\frac{1}{r+1}\right)}{3
   (r+1)^3}-\frac{16 \left(11 r^3+9 r^2-9 r-11\right) \text{Li}_2(-r)
   \ln (r+1)}{3 (r+1)^3}
\nn
&
 +\frac{8 \left(33 r^3+75 r^2+57 r+11\right)
   \text{Li}_2(-r) \ln (r)}{3 (r+1)^3}-16
   \text{Li}_4\left(\frac{1}{r+1}\right)-16
   \text{Li}_4\left(\frac{r}{r+1}\right)
\nn
&
  -8 \text{Li}_2(-r) \ln^2(r)+16 \text{Li}_2(-r) \ln (r) \ln (r+1)+8 \text{Li}_3(-r) \ln
   (r)+16 \text{Li}_3\left(\frac{1}{r+1}\right) \ln (r)
\nn
&
  +\frac{4
   \left(-40 \left(4 r^2 (27 \zeta (3)-2)+r (189 \zeta (3)-8)+99 \zeta
   (3)\right)+5 \pi^2 r \left(67 r^2+147 r+66\right)+33 \pi^4
   (r+1)^3\right)}{135 (r+1)^3}
\nn
&
    -\frac{32 \left(12 r^2+21 r+11\right)
   \ln^3(r+1)}{9 (r+1)^3}+\frac{4 r \left(11 r^2+21 r+12\right) \ln^3(r)}{9 (r+1)^3}-\frac{4}{3}
   \ln^3(r) \ln (r+1)
\nn
&
+\frac{16 \left(12 r^2+21 r+11\right) \ln (r) \ln^2(r+1)}{3 (r+1)^3}-\frac{4 r \left(67 r^2+141 r+60\right) \ln^2(r)}{9 (r+1)^3}
\nn
&
   +\frac{16 r \left(193 r^2+384 r+177\right) \ln
   (r)}{27 (r+1)^3}+\frac{4 \left(-11 r^3-9 r^2+9 r+11\right) \ln^2(r) \ln (r+1)}{3 (r+1)^3}
\nn
&
  -\frac{8 \left(\pi^2 \left(66 r^3+90
   r^2+9 r-33\right)+2 \left(193 r^3+561 r^2+561 r+193\right)\right)
   \ln (r+1)}{27 (r+1)^3}
\nn
&
  +\frac{8 \left(67 r^3+69 r^2-93 r+3 \pi^2
   (r+1)^3-67\right) \ln (r) \ln (r+1)}{9 (r+1)^3}-\frac{4}{3} \ln^4(r+1)
\nn
&
   +8 \ln^2(r) \ln^2(r+1)+16 \zeta (3) \ln
   (r+1)-16 \zeta (3) \log (r)+\frac{32 r \ln^2(r+1)}{3 (r+1)^2}.
\end{align}
%----------------------------------------------------------------------------------------------------
The following are the first 2 terms in the $\e$ expansion of $f_{n_f}$ from 
Eq.~(\ref{fnf:exp}).
%----------------------------------------------------------------------------------------------------
\begin{align}
f_{n_f}^{(0)}(r)
&= 
-\frac{16 \left(2 r (r+1)-2 (r+1)^3 \ln (r+1)+r (r (2 r+3)+3) \ln (r)\right)}{3 (r+1)^3},
\end{align}
%----------------------------------------------------------------------------------------------------
%----------------------------------------------------------------------------------------------------
\begin{align}
f_{n_f}^{(1)}(r)
&= \frac{8}{9 (r+1)^3}
\left(
  -12 \left(r^3-1\right)
    \text{Li}_2\left(-\frac{1}{r}\right)
  -32 r^3 \ln (r+1)
  +3 \pi ^2 r^2
  +20 r^2
\right.
\nn
& \quad \left.
  -96 r^2 \ln(r+1)
  -3 \left(4 r^3+3 r^2+3 r-2\right) \ln ^2(r)
  +4 \ln (r) 
    \left(
      3 \left(r^3-1\right) \ln (r+1)
\right.
\right.
\nn
& \quad \left. \left.
      +r \left(8 r^2+21r+3\right)
    \right)
  +3 \pi ^2 r
  +20 r
  -96 r \ln (r+1)
  -32 \ln (r+1)
  +2 \pi ^2
\right) .
\end{align}

\end{document}